\documentclass[aps,prd,twocolumn,superscriptaddress,nofootinbib]{revtex4-1}
\usepackage{amsfonts}
\usepackage{psfrag}
\usepackage{graphicx}
\usepackage{grffile}
\usepackage{cancel}
\usepackage{amsmath}
\usepackage{natbib}
\usepackage{xr}
\usepackage{hyperref}
\usepackage{cleveref}
\crefname{equation}{Eq.}{Eqs.}
\crefname{figure}{Fig.}{Figs.}

\usepackage{verbatim}
\usepackage{epstopdf}
\usepackage{subcaption}
\usepackage{multirow}
\usepackage{commath}
\usepackage{xcolor}
\usepackage{comment}
\hypersetup{    
  colorlinks      = true,
  linkcolor       = {blue},
  linkbordercolor = {white},
  citecolor       = {blue},
  citebordercolor = {white},
  urlcolor        = {blue},
  urlbordercolor  = {white},
}

\usepackage{soul}
\usepackage{xcolor}  % remove when done with large colored paragraphs

\usepackage{booktabs,dcolumn,caption}

\newcolumntype{d}[1]{D{.}{.}{#1}}

\RequirePackage[normalem]{ulem} %DIF PREAMBLE
\RequirePackage{color}\definecolor{RED}{rgb}{1,0,0}\definecolor{BLUE}{rgb}{0,0,1} %DIF PREAMBLE
\providecommand{\DIFaddbegin}{} %DIF PREAMBLE
\providecommand{\DIFaddend}{} %DIF PREAMBLE
\providecommand{\DIFdelbegin}{} %DIF PREAMBLE
\providecommand{\DIFdelend}{} %DIF PREAMBLE
\providecommand{\DIFaddbeginFL}{} %DIF PREAMBLE
\providecommand{\DIFaddendFL}{} %DIF PREAMBLE
\providecommand{\DIFdelbeginFL}{} %DIF PREAMBLE
\providecommand{\DIFdelendFL}{} %DIF PREAMBLE
\newcommand{\DIFscaledelfig}{0.5}
\RequirePackage{settobox} %DIF PREAMBLE
\RequirePackage{letltxmacro} %DIF PREAMBLE
\newsavebox{\DIFdelgraphicsbox} %DIF PREAMBLE
\newlength{\DIFdelgraphicswidth} %DIF PREAMBLE
\newlength{\DIFdelgraphicsheight} %DIF PREAMBLE
\LetLtxMacro{\DIFOincludegraphics}{\includegraphics} %DIF PREAMBLE
\newcommand{\DIFaddincludegraphics}[2][]{{\color{blue}\fbox{\DIFOincludegraphics[#1]{#2}}}} %DIF PREAMBLE
\newcommand{\DIFdelincludegraphics}[2][]{% %DIF PREAMBLE
\sbox{\DIFdelgraphicsbox}{\DIFOincludegraphics[#1]{#2}}% %DIF PREAMBLE
\settoboxwidth{\DIFdelgraphicswidth}{\DIFdelgraphicsbox} %DIF PREAMBLE
\settoboxtotalheight{\DIFdelgraphicsheight}{\DIFdelgraphicsbox} %DIF PREAMBLE
\scalebox{\DIFscaledelfig}{% %DIF PREAMBLE
\parbox[b]{\DIFdelgraphicswidth}{\usebox{\DIFdelgraphicsbox}\\[-\baselineskip] \rule{\DIFdelgraphicswidth}{0em}}\llap{\resizebox{\DIFdelgraphicswidth}{\DIFdelgraphicsheight}{% %DIF PREAMBLE
\setlength{\unitlength}{\DIFdelgraphicswidth}% %DIF PREAMBLE
\begin{picture}(1,1)% %DIF PREAMBLE
\thicklines\linethickness{2pt} %DIF PREAMBLE
{\color[rgb]{1,0,0}\put(0,0){\framebox(1,1){}}}% %DIF PREAMBLE
{\color[rgb]{1,0,0}\put(0,0){\line( 1,1){1}}}% %DIF PREAMBLE
{\color[rgb]{1,0,0}\put(0,1){\line(1,-1){1}}}% %DIF PREAMBLE
\end{picture}% %DIF PREAMBLE
}\hspace*{3pt}}} %DIF PREAMBLE
} %DIF PREAMBLE
\LetLtxMacro{\DIFOaddbegin}{\DIFaddbegin} %DIF PREAMBLE
\LetLtxMacro{\DIFOaddend}{\DIFaddend} %DIF PREAMBLE
\LetLtxMacro{\DIFOdelbegin}{\DIFdelbegin} %DIF PREAMBLE
\LetLtxMacro{\DIFOdelend}{\DIFdelend} %DIF PREAMBLE
\DeclareRobustCommand{\DIFaddbegin}{\DIFOaddbegin \let\includegraphics\DIFaddincludegraphics} %DIF PREAMBLE
\DeclareRobustCommand{\DIFaddend}{\DIFOaddend \let\includegraphics\DIFOincludegraphics} %DIF PREAMBLE
\DeclareRobustCommand{\DIFdelbegin}{\DIFOdelbegin \let\includegraphics\DIFdelincludegraphics} %DIF PREAMBLE
\DeclareRobustCommand{\DIFdelend}{\DIFOaddend \let\includegraphics\DIFOincludegraphics} %DIF PREAMBLE
\LetLtxMacro{\DIFOaddbeginFL}{\DIFaddbeginFL} %DIF PREAMBLE
\LetLtxMacro{\DIFOaddendFL}{\DIFaddendFL} %DIF PREAMBLE
\LetLtxMacro{\DIFOdelbeginFL}{\DIFdelbeginFL} %DIF PREAMBLE
\LetLtxMacro{\DIFOdelendFL}{\DIFdelendFL} %DIF PREAMBLE
\DeclareRobustCommand{\DIFaddbeginFL}{\DIFOaddbeginFL \let\includegraphics\DIFaddincludegraphics} %DIF PREAMBLE
\DeclareRobustCommand{\DIFaddendFL}{\DIFOaddendFL \let\includegraphics\DIFOincludegraphics} %DIF PREAMBLE
\DeclareRobustCommand{\DIFdelbeginFL}{\DIFOdelbeginFL \let\includegraphics\DIFdelincludegraphics} %DIF PREAMBLE
\DeclareRobustCommand{\DIFdelendFL}{\DIFOaddendFL \let\includegraphics\DIFOincludegraphics} %DIF PREAMBLE
\begin{document}

\graphicspath{ {images/} }

\title{Dynamical friction in self-interacting ultralight dark matter}

\date{\today}

\author{Noah Glennon}
\email{nglennon@wildcats.unh.edu}
\affiliation{Department of Physics and Astronomy, University of New Hampshire, Durham, New Hampshire 03824, USA}

\author{Nathan Musoke}
\email{nathan.musoke@unh.edu}
\affiliation{Department of Physics and Astronomy, University of New Hampshire, Durham, New Hampshire 03824, USA}

\author{Ethan~O.~Nadler}
\email{enadler@carnegiescience.edu}
\affiliation{Carnegie Observatories, 813 Santa Barbara Street, Pasadena, CA 91101, USA}
\affiliation{Department of Physics $\&$ Astronomy, University of Southern California, Los Angeles, CA, 90007, USA}

\author{Chanda Prescod-Weinstein}
\email{chanda.prescod-weinstein@unh.edu}
\affiliation{Department of Physics and Astronomy, University of New Hampshire, Durham, New Hampshire 03824, USA}

\author{Risa~H.~Wechsler}
\email{rwechsler@stanford.edu}
\affiliation{Kavli Institute for Particle Astrophysics and Cosmology and Department of Physics, Stanford University, Stanford, CA 94305, USA}
\affiliation{SLAC National Accelerator Laboratory, Menlo Park, CA 94025, USA}

\begin{abstract}
We explore how dynamical friction in an ultralight dark matter (ULDM) background is affected by dark matter self-interactions.  We calculate the force of dynamical friction on a point mass moving through a uniform ULDM background with self-interactions, finding that the force of dynamical friction vanishes for sufficiently strong repulsive self-interactions. Using the pseudospectral solver $\texttt{UltraDark.jl}$, we show with simulations that reasonable values of the ULDM self-interaction strength and particle mass cause $\mathcal{O}(1)$ differences in the acceleration of an object like a supermassive black hole (SMBH) traveling near the center of a soliton, relative to the case with no self-interactions.  For example, repulsive self-interactions with $\lambda = 10^{-90}$ yield a deceleration due to dynamical friction $\approx70\%$ smaller than a model with no self-interactions. We discuss the observational implications of our results for SMBHs near soliton centers and for massive satellite galaxies falling into ultralight axion halos and show that outcomes are dependent on whether a self-interaction is present or not.
\end{abstract}

\maketitle

\section{Introduction}
\label{sec:intro}

Cosmological observations indicate that most of the matter in the Universe is dark.  Such evidence comes, for example, from the cosmic microwave background (CMB), galactic rotation curves, and gravitational lensing~\cite{Ade2015, Bertone2005, Bertone2018, Freese2009, Chabanier:2019eai}.  Many dark matter models fit into the cold dark matter (CDM) paradigm, which is characterized by non-relativistic  and collisionless dark matter particles~\cite{Buckley2018, Armendariz2014}.  While CDM models have been very successful at predicting the large-scale structure of the universe, little is known about the microphysics of dark matter.

An alternative to traditional CDM models is ultralight dark matter (ULDM). In ULDM models, the constituent particles have masses around $10^{-22}~\rm{eV}$, and therefore have a de Broglie wavelength on the order of a kiloparsec.  While they behave very similarly to CDM on large scales, ULDM models can yield distinct predictions on small scales due to the wave-like nature of the dark matter particles, which may alleviate potential tensions between CDM and dwarf galaxy observations~(e.g., see \cite{Bullock:2017xww} for a review). For example, the ``cusp--core'' problem concerns CDM predictions for dense cusps at the center of dark matter halos, which may differ from roughly constant-density cores observed in some dwarf galaxies. Although core formation can largely be explained by baryonic processes (e.g., see Refs.~\cite{Popolo2016, Marsh2015}), ULDM may provide an alternative solution. In particular, because the de Broglie wavelength is very large, structure on scales smaller than this wavelength is naturally smoothed out.

ULDM is motivated by specific particle physics models such as axion-like particles (ALPs) resulting from string theory. 
ALPs may have a large range of possible masses from $1\rm{eV}$ down to $10^{-33}~\rm{eV}$~\cite{Arvanitaki2010, Marsh2016} and ultralight axions (ULAs) with mass $m \lesssim 10^{-18}\;\mathrm{eV}$ are a subclass of these models~\cite{Hui2017}. They are not the same as the QCD axions that were theorized to solve the strong CP problem in QCD~\cite{Peccei1977}. In this paper, we focus on a general class of ULDM models that encompasses ULAs. Crucially, we consider models that feature self-interactions between dark matter particles, which are a generic feature of many ULDM scenarios~\cite{Mehta:2021pwf}.

In many ULDM models, self interactions are neglected because they are constrained to be very small~\cite{Sikivie:2009fv, Kirkpatrick:2021wwz, Fan2016, DES:2020fxi, Chavanis2016, Cembranos:2018ulm, Dave:2023wjq, Delgado:2022vnt, Li:2013nal}.  However, as shown in, e.g., Refs.~\cite{Chavanis2016, Glennon2020, Glennon2022, Desjacques2017, Rindler2011, Dmitriev:2021utv}, even small self-interactions can be important for understanding structure formation in ULA models.  This is because the non-linear effects from self-interactions are determined by the self-coupling times the phase space density of the dark matter particle in the environment (which may be very large) rather than just the strength of the self-coupling~\cite{Desjacques2017}.  Predicted differences depend on the sign of the self-interaction, and include phenomena like solitonic collapse, oscillations, and explosions when there are attractive self-interactions~\cite{Chavanis2016, Glennon2020}, and vortex formation when there are repulsive self-interactions~\cite{Rindler2011,Dmitriev:2021utv}.  Self-interactions have also been shown to affect the tidal disruption timescales of solitons undergoing tidal stripping~\cite{Glennon2022}.

It is interesting to consider how the motion of an object traveling through an ULDM background is affected by the ultra-light nature of the surrounding particles and how this motion is affected by the presence of self-interactions. For example, dynamical friction is the process by which an object is slowed down by the gravitational interactions with the matter around it \cite{Chandrasekhar1943}.  When  a massive object moves through a background medium, the particles in the medium will be accelerated by the massive object's gravity. In a standard picture, the particles then form an overdense region trailing the massive object, referred to as a gravitational ``wake,'' and slow the massive object down through gravitational interactions.
Dynamical friction is important, for example, to understand why heavier galaxies tend to be found near the centers of galaxy clusters~\cite{vandenBosch:1998dc, Fujii:2005kw} and why supermassive black holes (SMBHs) migrate towards the centers of galaxies~\cite{Antonini2012}.

Dynamical friction in an ULDM background differs from that in traditional CDM models.  In particular, due to ULAs' wave-like properties, gravitational wakes are suppressed.  Since the overdense region behind the massive object is smaller than what one would expect in CDM models, there will be a smaller dynamical friction force on the object~\cite{Hui2017}.  The effects of dynamical friction have been studied in detail for ULDM models without self-interactions~\cite{Lancaster:2019mde, Wang:2021udl, Bar2019ApJ, Buehler:2022tmr}.

ULDM self-interactions may affect dynamical friction. For example, in the presence of attractive self-interactions, overdense gravitational wakes will be denser than they would be if there were no self-interactions.  Conversely, repulsive self-interactions are expected to reduce the size of density wakes. In this work, we therefore calculate how self-interactions change the dynamical friction force on an object traveling through an ULDM background.

The paper is organized as follows.  In Sec.~\ref{sec:ULA}, we describe the ULDM model we use and summarize the current constraints on quartic self-interactions. Sec.~\ref{sec:description} describes the code used for our numerical simulations, \texttt{UltraDark.jl}, and describes the simulations' astrophysical context.  We present the results of our simulations in Sec.~\ref{sec:results}.  These include an analysis of theoretical predictions from our analytic calculation, where applicable, and an analysis of simulations corresponding to a realistic scenario of an SMBH traveling through an ULDM background with self-interactions.  We summarize our results and discuss their implications in Sec.~\ref{sec:conclusions}.

\section{Ultralight Dark Matter Model and Physical Setup}
\label{sec:ULA}

\subsection{ULDM Model}

Herein we assume that the ULDM can be treated as a classical field that is minimally coupled to gravity.  From this assumption, the action becomes
\begin{equation}
    S = \int d^4 x \sqrt{-g} \left[\frac{1}{2} g^{\mu\nu} \partial_{\mu}\phi\partial_{\nu}\phi-\frac{1}{2}m^2\phi^2-\frac{\lambda}{4}\phi^4\right]
\end{equation}
where $\phi$ is the scalar field, $m$ is the mass of the field, and $\lambda$ is the dimensionless self-coupling.  This action uses only the leading order self-interaction term.  By writing the real scalar field $\phi$ in terms of a complex field $\psi$,
\begin{equation}
    \phi = \frac{\hbar}{\sqrt{2m}}\left(\psi e^{-i m t/\hbar} + \psi^* e^{i m t/\hbar}\right)
\end{equation}
we arrive at the equations of motion.  The equations of motion in the Newtonian gauge are the Gross-Pitaevskii-Poisson (GPP) equations, 
\begin{equation}
	i \hbar \dot{\psi} = -\frac{\hbar^2}{2m} \nabla^2 \psi + m \Phi \psi + \frac{\hbar^3 \lambda}{2m^3} \abs{\psi}^2 \psi
	\label{eq:GPeqn}
\end{equation}
and
\begin{equation}
	\nabla^2 \Phi = 4\pi G m \abs{\psi}^2.
	\label{eq:Peqn}
\end{equation}
Here, $\Phi$ is the gravitational potential and $\psi$ is the ULDM field (see Ref.~\cite{Kirkpatrick2020} for a more detailed derivation).  We assume a particle mass of $m=10^{-22}~\mathrm{eV}$ and a dimensionless coupling $\kappa$ related to $\lambda$ by
\begin{equation}
    \kappa = 2.1 \times 10^{88}\lambda.
    \label{eq:kaptolambda}
\end{equation}

\subsection{Constraints on ULDM Self-interactions}
\label{sec:constraints}

To contextualize our choice of self-interaction strengths below, we summarize constraints on this term from the literature. For \emph{attractive} self-interactions, Refs.~~\cite{Sikivie:2009fv, Kirkpatrick:2021wwz} predict that the self-interaction strength scales as
\begin{equation}
    \lambda \approx -\frac{m^2}{f^2},
\end{equation}
where $f$ is the axion decay constant.  This follows from expanding the axion potential
\begin{equation}
    V(\phi) = m^2 f^2 \left( 1-\mathrm{cos}(\phi/f) \right).
\end{equation}
Assuming values of $m = 10^{-22}~\text{eV}$ and $f=10^{17}~\text{GeV}$, this leads to a predicted self-coupling of $\lambda \approx -1 \times 10^{-96}$ or $\kappa \approx -2\times10^{-8}$.  This is several orders of magnitude smaller than the typical self-interaction strengths we will adopt here, which are similar to those assumed in previous studies of ULDM self-interactions (e.g.~\cite{Glennon2022}).

For \emph{repulsive} self-interactions, Ref.~\cite{Fan2016} presents a rough estimate for the strength of repulsive self-interactions allowed in ULDM models.  This estimate is based on the fact that the linear matter-power spectrum is well constrained on large scales.  Thus, the Jeans scale induced by the repulsive self-interactions must be less than $\sim 1~\text{Mpc}$.  This leads to an estimated constraint on the quartic interaction that depends on the boson mass and axion decay constant via
\begin{equation}
    \begin{split}
        k_{\lambda}^{\text{eq}} = 2.7 \left(\frac{m}{10^{-20}~\text{eV}}\right)\left(\frac{f}{10^{13}~\text{Gev}}\right)
        \\
        \times \sqrt{\frac{1}{\lambda^{\text{eff}}_4}}\left(\frac{a}{a_{\text{eq}}}\right)~\text{Mpc}^{-1} \gtrsim 1~\text{Mpc}^{-1},
    \end{split}
\end{equation}
where $\lambda^{\text{eff}}_4 = \frac{1}{4!} \lambda \left(\frac{m}{f}\right)^2$.  Using a mass of $m=10^{-22}~\text{eV}$ and decay constant of $10^{17}~\text{GeV}$, the constraint is approximately $\lambda < 3\times10^{-93}$.\footnote{We expect that smaller-scale data is even more sensitive to such self-interactions; this is an interesting area for future study.} Again, this is slightly smaller than the typical self-interaction strengths we will assume. 

There are also other estimates of allowed self-interaction strengths; we note that these estimates often differ by orders of magnitude. For example, Ref.~\cite{Chavanis2016} use observations of the Bullet Cluster and of one of the smallest known galaxies, Willman I, to claim that the dimensionless coupling is $\lambda = 3.7\times10^{-14}$ for repulsive self-interactions and $\lambda = -2.0\times10^{-86}$ for attractive self-interactions.  However, these values of $\lambda$ assume different ULDM masses than our fiducial mass of $10^{-22}~\text{eV}$.  In particular, the masses found for the attractive and repulsive cases are $m = 2.57\times10^{-20}~\text{eV}$ and $m = 1.69\times10^{-2}~\text{eV}$ respectively and the corresponding scattering lengths are $a_s = -8.29\times10^{-60}~\text{fm}$ $a_s = 1.73\times10^{-5}~\text{fm}$ respectively.  The dimensionless coupling corresponding to these values is
\begin{equation}
    \kappa = \frac{4\pi \hbar a_s}{\mathcal{T}m^2 G},
\end{equation}
yielding $\kappa = -3.3\times10^{-5}$ for attractive self-interactions and $\kappa=1.6\times10^{14}$ for repulsive self-interactions. These values are consistent with the stronger constraints from Ref.~\cite{Fan2016}.

Meanwhile, according to Ref.~\cite{Li:2013nal}, the limits on the dimensionful couplings are
\begin{equation}
    9.5\times10^{-19}~\frac{\text{cm}^3}{\text{eV}} < \lambda < 4\times10^{-17}~\frac{\text{cm}^3}{\text{eV}}.
\end{equation}
These limits are based on observations from the CMB and the number of light particle species produced at matter--radiation equality, $z_{\text{eq}}$.  In particular, the upper limits are needed so that ULDM behaves like CDM slightly before $z_{\text{eq}}$; the lower bounds are determined using constraints from the effective number of relativistic degrees of freedom $N_{\text{eff}}$.  The dimensionless coupling is related to the dimensionful coupling by
\begin{equation}
    \hat{\lambda} = \lambda\frac{m^2 c}{\hbar^3}.
\end{equation}
Assuming a boson mass of $m = 10^{-22}~\frac{\text{eV}}{c^2}$, the constraint on the dimensionless coupling becomes
\begin{equation}
    5.2\times10^{-91} < \hat{\lambda} < 1.2\times10^{-90},
\end{equation}
which implies
\begin{equation}
    0.00026 < \kappa < 0.0108.
\end{equation} 
This constraint arises because, at sufficiently early times, the quartic self-interaction term causes the bosons to behave like radiation.  However, at Big Bang Nucleosynthesis, the temperature is such that Hubble friction makes the bosons behave like dark energy. Similarly, Ref.~\cite{Cembranos:2018ulm} claims that, in the presence of repulsive self-interactions, ULDM behaves like radiation at early times.  This effect can be used to constrain the strength of the self-coupling, yielding
\begin{equation}
    \log_{10}(\lambda) < -91.86 + 4\log_{10}\left(\frac{m}{10^{-22}~\rm{eV}}\right)
\end{equation}
for particle masses larger than $10^{-24}~\rm{eV}$.  For a particle mass of $10^{-22}~\rm{eV}$, this corresponds to $\kappa < 2.9$.

Finally, in Ref.~\cite{Dave:2023wjq}, the authors constrain ULDM self-interactions using the soliton--halo relation, which includes self-interactions, and fitting galactic rotation curves using scaled solutions of the GPP equations. The galactic rotation curves studied by Ref.~\cite{Dave:2023wjq} can be fit by assuming repulsive self-interactions with $\lambda\gtrsim\mathcal{O}(10^{-90})$ or $\kappa\gtrsim\mathcal{O}(10^{-2})$. Ref.~\cite{Delgado:2022vnt} presents a similar analysis, resulting in a best-fit particle mass of $m=2.2\times10^{-22}~\mathrm{eV}$ and scattering length of $a_s = 7.8\times10^{-62}~\mathrm{fm}$, which corresponds to a self-interaction strength of $\lambda = 2.2\times10^{-90}$ ($\kappa = 4.2\times10^{-3}$). Note that Eq.~\ref{eq:kaptolambda} is not used in this translation because the particle mass is not assumed to be $m=10^{-22}~\mathrm{eV}$.

\subsection{Dynamical Friction Expectations}

The classic Chandrasekhar formulation of dynamical friction \cite{Chandrasekhar1943} describes the force on an object from gravitational interactions with stars.  This was done assuming the field of stars was collisionless and neglecting the self-gravity of the stars.  In this scenario, the dynamical friction force for an object of mass $M$ traveling through a background of particles with mass $m \ll M$ is given by
\begin{equation}
    F_{\mathrm{DF}} = -4\pi G^2 M^2 \frac{\bf{v_{\mathrm{rel}}}}{v_{\mathrm{rel}}^3}\rho(<v_{\mathrm{rel}})\ln{(\Lambda)}
\end{equation}
where $\bf{v_{\mathrm{rel}}}$ is the massive object's velocity relative to the background, ${v_{\mathrm{rel}}} = \abs{\bf{v_{\mathrm{rel}}}}$, $\rho(<v_{\mathrm{rel}})$ is the background density of stars with velocity less than $v_{\mathrm{rel}}$, and $\ln \Lambda$ is the Coulomb logarithm given by $\ln{\Lambda} = \ln({b_{\mathrm{max}}/b_{\mathrm{min}}})$, where $b_{\mathrm{max}}$ and $b_{\mathrm{min}}$ are the maximum and minimum impact parameters in the system~\cite{2008gady.book.....B, Chandrasekhar1943}. 

This description of dynamical friction does not apply in many situations because the dynamical friction force is proportional to the local matter density. For example, if an object exists outside of a halo where the local density vanishes, the object should still feel a dynamical friction force~\cite{2021ApJ...912...43B}.  Meanwhile, if an object falls towards the center of a constant density core described by a harmonic potential, the infalling object will stall and feel no dynamical friction force~\cite{2006MNRAS.373.1451R}. While many formulations of dynamical friction assume a homogeneous background and that dynamical friction is a local phenomena, Ref.~\cite{1984MNRAS.209..729T} develops a perturbative description without assuming a uniform density background.  They find the dynamical friction force is only due to torques applied by particles in resonant orbits with the perturber, known as LBK torques, as first described in Ref.~\cite{1972MNRAS.157....1L}.  This non-local description resolves the core-stalling issue that arises in the canonical description of dynamical friction.

In this work, we perform a perturbative calculation to predict the dynamical friction force on a point mass in a uniform ULDM background with self interactions.  This calculation follows a similar procedure found in Ref.~\cite{Lancaster:2019mde}, which derives the dynamical friction force on a point mass moving through a uniform ULDM background without self-interactions.  

First, we write 
\begin{equation}
    \mathbf{u} = \frac{\hbar}{m}\nabla \theta,
\end{equation}
where $\theta$ is dark matter phase.  We also define
\begin{equation}
    \psi = \sqrt{\rho} e^{i\theta}.
\end{equation}
Rewriting the GPP equations in terms of two real partial differential equation, we obtain
\begin{equation}
    \frac{\partial\rho}{\partial t} + \nabla \cdot (\rho \mathbf{u}) = 0
\end{equation}
\begin{equation}
    \frac{\partial \mathbf{u}}{\partial t} + (\mathbf{u} \cdot \nabla) \mathbf{u} = -\nabla U - \nabla U_{Q},
\end{equation}
where
\begin{equation}
    U_Q = -\frac{\hbar}{2m^2}\frac{\nabla^2 \sqrt{\rho}}{\sqrt{\rho}}.
\end{equation}
By replacing $\rho \rightarrow \bar{\rho} + \delta \rho$ and $\bf{u} \rightarrow \bf{v}_{\rm{rel}} + \delta \bf{v}$, we arrive at Equations 4.8 and 4.9 in Ref.~\cite{Lancaster:2019mde},
\begin{equation}
        \frac{\partial \alpha}{\partial t} + (v_{\mathrm{rel}}\cdot \nabla) \alpha +\nabla \cdot \delta v = 0
\end{equation}
and
\begin{equation}
    \frac{\partial \delta v}{\partial t} + (v_{\mathrm{rel}}\cdot \nabla) \delta v = - \nabla U + \frac{\hbar^2}{4m^2} \nabla (\nabla^2 \alpha).
\end{equation}
Here, $\alpha(x) \equiv (\rho(x) - \bar{\rho})/\bar{\rho}$ and $v_{\text{rel}}$ is the speed of the point mass relative to the background. 

By combining the first two equations and assuming $v_{\mathrm{rel}}$ is in the $z$ direction, we obtain
\begin{equation}
    \frac{\partial^2 \alpha}{\partial t^2} - 2 v_{\mathrm{rel}}\frac{\partial^2 \alpha}{\partial z \partial t} + v_{\mathrm{rel}}^2 \frac{\partial^2 \alpha}{\partial z^2} - \nabla^2 U + \frac{\hbar^2}{4 m^2}\nabla^4 \alpha = 0.
\end{equation}
We then write $U$ as
\begin{equation}
    U = -\frac{GM}{r} + \frac{\hbar^3 \lambda}{2 m^4}\bar{\rho} \alpha.
\end{equation}
Assuming time independence and substituting in for $U$, we obtain
\begin{equation}
    v_{\mathrm{rel}}^2\frac{\partial^2 \alpha}{\partial z^2} + \frac{\hbar^2}{4m^2}\nabla^4\alpha = 4\pi G M \delta^3(r) + \frac{\hbar^3 \lambda}{2m^4}\bar{\rho}\nabla^2 \alpha.
\end{equation}
Next, we perform a Fourier transform to derive
\begin{equation}
    \left(\frac{\hbar^2}{4m^4}k^4-v_{\mathrm{rel}}^2 k_z^2 + \frac{\hbar^3 \lambda}{2m^4}\bar{\rho}k^2\right)\hat{\alpha} = 4\pi GM,
\end{equation}
where $\hat{\alpha}$ is the Fourier transform of $\alpha$. We define several new variables to simplify this equation with $\tilde{\kappa} \equiv \lambdabar k$, $\lambdabar \equiv \hbar/mv_{\mathrm{rel}}$, $M_{\mathrm{Q}} \equiv v_{\mathrm{rel}}/V_{\mathrm{Q}}$, $V_Q \equiv \hbar/m L_{\mathrm{Q}}$, $L_{\mathrm{Q}} \equiv \hbar^2/G Mm^2$, and $\tilde{r} \equiv r/\lambdabar$. With these definitions, we obtain
\begin{equation}
    \hat{\alpha} = \frac{16\pi \lambdabar^3}{M_{\mathrm{Q}}(\tilde{k}^4-4\tilde{k}^2_z+\xi \tilde{k}^2)}.
\end{equation}
Here, we have defined the dimensionless parameter $\xi$ as
\begin{equation}
    \xi = \frac{2\hbar^3\lambda}{m^4 v^2_{\text{rel}}}\bar{\rho} = 4 \left(\frac{c_\text{s}}{v_{\text{rel}}}\right)^2,
    \label{xi}
\end{equation}
where $c_s$ is the sound speed in the ULDM medium~\cite{Chavanis:2020rdo}. We parameterize the dynamical friction force for a given simulation setup using $\xi$.

We attempt to analytically calculate the dynamical friction force in the presence of ULDM self-interactions in Appendix~\ref{sec:analyticcalc}. Our perturbative calculation breaks down at intermediate self-interaction strengths ($1\lesssim \xi< 4$), so we do not compare our simulations to it in this regime, and we leave a treatment of this issue to future work. Nonetheless, this calculation clearly predicts that the dynamical friction force \emph{vanishes} for sufficiently repulsive self-interactions (i.e., $\xi>4$). In this limit, the sound speed in the ULDM medium exceeds relative velocities, preventing the formation of gravitational wakes; this is consistent with our simulations and provides a physical interpretation of our results. On the other hand, as self-interactions are turned off ($\xi\rightarrow 0$), the dynamical friction force predicted by our analytic calculation approaches the case without self-interactions. As we will show, this intuitive result also agrees with our simulations.

\section{Simulations}
\label{sec:description}

\subsection{Numerical Methods}

We perform simulations using an extension of \texttt{UltraDark.jl}, a pseudo-spectral solver for the GPP equations~\cite{ultradark}.
\texttt{UltraDark.jl} has been used in related work to investigate tidal stripping with ULDM self-interactions, vortices, and systems with several different ULDM species~\cite{Glennon2022,Glennon:2023oqa,Glennon:2023jsp}.

\texttt{UltraDark.jl} uses code units internally.
In these units, time, length, mass and self-interaction strength $\lambda$ are scaled by, respectively,
\begin{gather}
    \label{eq:time_scale}
    \mathcal{T}
    =
    {\left(
        \frac{3}{8 \pi} H_0^2 \Omega_{m, 0}
    \right)}^{-1/2}
    \approx
    74 \;\mathrm{~Gyr}
    \\
    \begin{split}
        \label{eq:length_scale}
        \mathcal{L}
        &=
        {\left(
                \frac{\hbar}{m}
        \right)}^{1/2}
        {\left(
                \frac{3}{8\pi}
                \Omega_{m, 0} H_0^2
        \right)}^{-1/4}
        \\
        &\approx
        38 {\left(\frac{10^{-22}\mathrm{~eV}}{m}\right)}^{1/2} \;\mathrm{~kpc}
    \end{split}
    \\
    \begin{split}
        \label{eq:mass_scale}
        \mathcal{M}
        &=
        {\left(
                \frac{\hbar}{m}
        \right)}^{3/2}
        \frac{1}{G}
        {\left(
                \frac{3}{8\pi} \Omega_{m, 0} H_0^2
        \right)}^{1/4}
        \\
        &\approx
        2.2 \times 10^{6} {\left(\frac{10^{-22}\mathrm{~eV}}{m}\right)}^{3/2} \;M_{\odot}
    \end{split}
    \\
    \kappa
    = \frac{\hbar^2}{2 \mathcal{T} m^3 G c} \lambda
    \approx 2.1 \times 10^{88} \lambda
    \;.
\end{gather}

The approximate values are computed with $H_0 = 70 \;\mathrm{~km/s/Mpc}$ and $\Omega_{m,0} = 0.3$. Units are chosen such that Eq.~\ref{eq:GPeqn} is transformed to a dimensionless form
\begin{gather}
    i \frac{\partial\psi'}{\partial t'}
    =
    - \frac{1}{2} \nabla'^{2} \psi'
    + \psi' \left(\Phi'(\mathbf{r}') + \Phi_{\text{particle}}'(\mathbf{r}) \right) + \kappa |\psi|^2 \psi
    \\
    \nabla'^{2} \Phi'(\mathbf{r})
    =
    4 \pi |\psi'|^{2}
    \;.
\end{gather}

\begin{figure*}[t!]
    \centering
    \includegraphics[width=.48\textwidth]{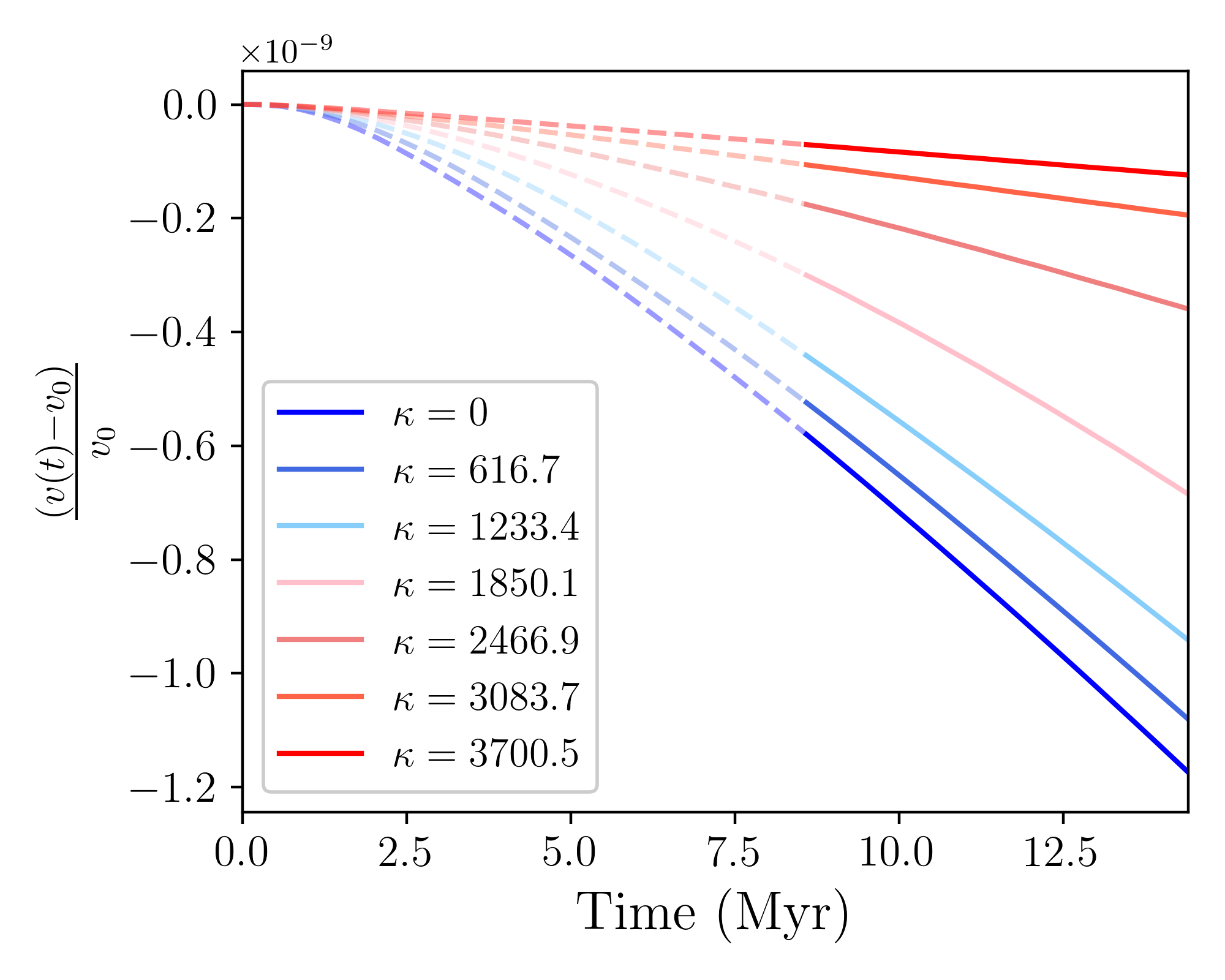}
    \includegraphics[width=.48\textwidth]{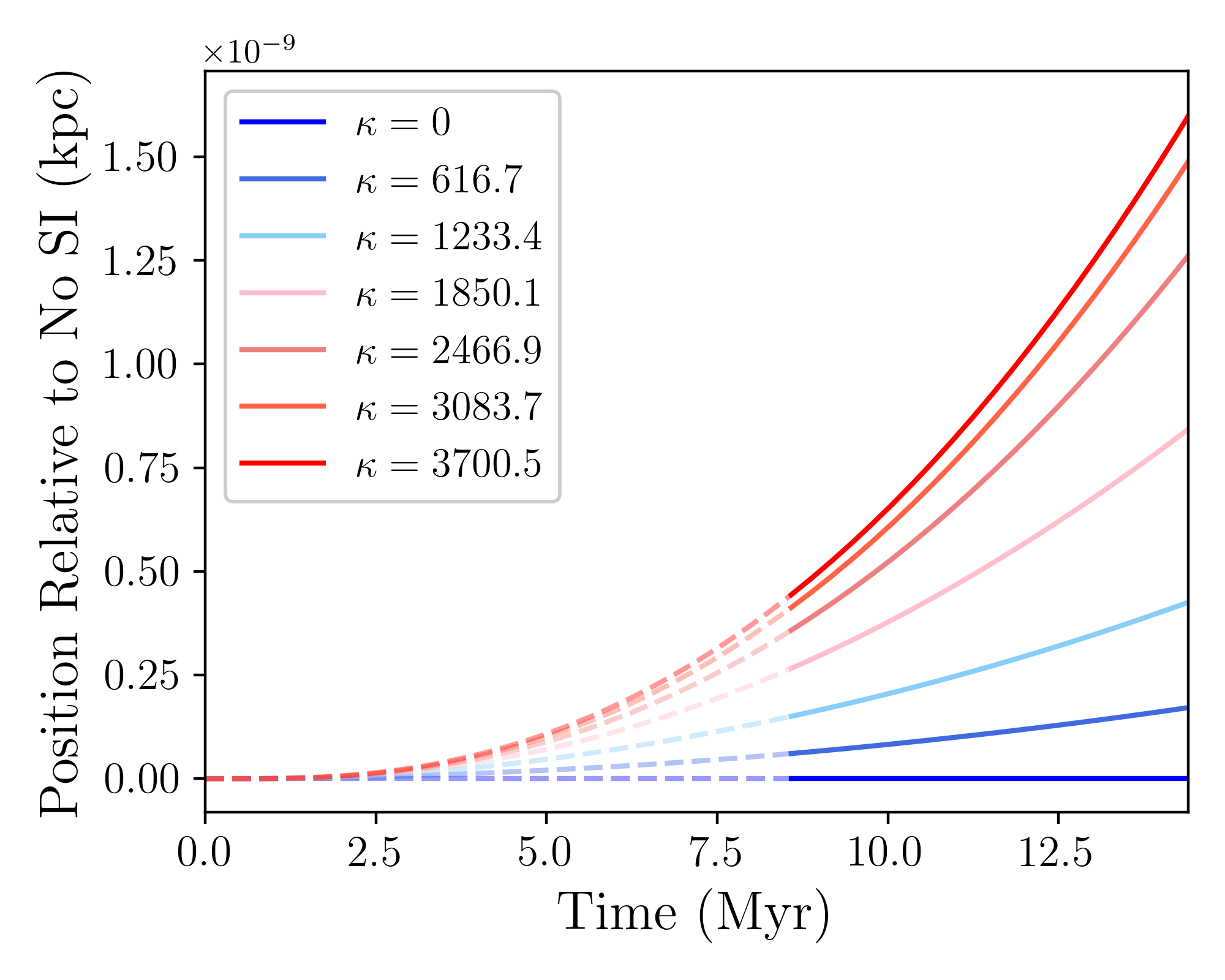}
    \caption{Left panel:  fractional change in velocity as a function of time in simulations of a point mass traveling through a uniform ULDM background.  The solid lines represent the portion of the simulation where we assume the point mass is approximately in a steady state to calculate the deceleration due to dynamical friction. Right panel: same as the left panel, but showing the difference in  position of the point mass. Various repulsive self-interaction strengths, shown by colored lines, are compared to the position of the same point mass with no self-interactions (dark blue). Repulsive self-interactions reduce the dynamical friction force, causing the point mass to move further in a given amount of time compared to the case without self-interactions.
    }
    \label{velvstime}
\end{figure*}

The gravitational potential $\Phi_{\text{particle}}$ is that of a particle that interacts gravitationally with the ULDM field.
It is pre-computed and shifted to the position of the particle.
We added two classes of particles to \texttt{UltraDark.jl}: point particles with potential
\begin{equation}
    \label{eq:point_potential}
    \Phi_{\text{point}} = -\frac{G M}{r}
\end{equation}
and Plummer spheres with potential 
\begin{equation}
\label{eq:plummer_potential}
    \Phi_{\text{Plummer}} = -\frac{G M}{\sqrt{r^2 + a^2}}
\end{equation}
where $M$ is the mass of the point particle or Plummer sphere and $a$ is the Plummer radius.
Note that the potential in \cref{eq:point_potential} is singular and can lead to numerical overflow if the location of the particle coincides with a grid point. We avoid this by simulating particles that travel along straight lines, coming no less than half a grid spacing from these formally singular values. Throughout, we compare our results with a point particle potential to simulations that use the Plummer sphere potential in \cref{eq:plummer_potential}, which smooths out this divergence and is commonly  gravitational softening.

The gravitational potential of the ALP field is calculated in Fourier space, 
\begin{equation}
    \Phi(\mathbf{x}, t + h)
    =
    \mathcal{F}^{-1} \left\{
        -\frac{4 \pi}{k^2} \mathcal{F}\left\{ |\psi|^2 \right\}
    \right\}
\end{equation}
where $\mathbf{k}$ is the coordinate in Fourier space.
Note that the gravitational potential of the $\Phi$ is periodic while $\Phi_{\text{particle}}$ is not.

For each time step with size $h$, the ULDM field and particle position are updated by a symmetrized split step method.
The code first updates the velocities of the ULDM field and particle by a half time step $h/2$,
\begin{gather}
    \label{eq:update_phase_1}
    \psi \to \exp\left( - i\frac{h}{2} \left(\Phi + \Phi_{\text{particle}}\right) \right) \exp\left(-i \frac{h}{2} \kappa |\psi|^2 \right) \psi
    \\
    \label{eq:update_velocity_1}
    \mathbf{v} \to \mathbf{v} + \frac{h}{2} \times \left[\mathcal{F}^{-1}\left\{ -i \mathbf{k} \mathcal{F}\left\{\Phi\right\} \right\}\right]_{\mathbf{x}}
    \;.
\end{gather}
We then perform a whole-step update to the ULDM density and particle position based on these velocities,
\begin{gather}
    \label{eq:update_density}
    \psi \to \mathcal{F}^{-1}\left\{ \exp\left(-i h \frac{k^2}{2} \right) \mathcal{F}\left\{ \psi \right\} \right\}
    \\
    \label{eq:update_position}
    \mathbf{x} \to \mathbf{x} + h \mathbf{v}
    \;.
\end{gather}
Finally, the velocities of the ULDM field and particle are updated by the remaining half time step $h/2$,
\begin{gather}
    \label{eq:update_phase_2}
    \psi \to \exp\left( - i\frac{h}{2} \left(\Phi + \Phi_{\text{particle}}\right) \right) \exp\left(-i \frac{h}{2} \kappa |\psi|^2 \right) \psi
    \\
    \label{eq:update_velocity_2}
    \mathbf{v} \to \mathbf{v} + \frac{h}{2} \times \left[\mathcal{F}^{-1}\left\{ -i \mathbf{k} \mathcal{F}\left\{\Phi\right\} \right\}\right]_{\mathbf{x}}
    \;.
\end{gather}
In each of these equations, $\mathbf{x}$ and $\mathbf{v}$ are the position and velocity of the particle, and $\mathcal{F}$ is the Fourier transform.
The notation $[f]_{\mathbf{x}}$ indicates the component of a matrix $f$ at the grid cell containing position $\mathbf{x}$.

The time step $h$ is chosen to be small enough such that the particle does not cross more than 1/4 of a grid cell in each step, and such that the change in the phase due to \cref{eq:update_velocity_1,eq:update_velocity_2} does not exceed $2\pi$. This can be expressed as
\begin{equation}
    \label{eq:timestep}
    h
    \leq
    \min\left(
        \frac{l_{\text{box}}}{\texttt{resol}} \frac{1}{4 |\mathbf{v}|},
        \frac{2 \pi}{\max(\Phi_{\text{total}})},
        \frac{2 \pi}{\max(k)}
    \right)
    \;,
\end{equation}
where \texttt{resol} is the resolution of the grid and maxima of $\Phi$ and $k$ are computed over the entire box. Time stepping is adaptive and time step sizes update after every 10 steps.
Note that because the first and last steps are identical and only update velocities (and do not depend on the velocities themselves), they can be combined if consecutive steps have the same size.

Because \texttt{UltraDark.jl} uses FFTs, the ULDM field has periodic boundary conditions topologically equivalent to a torus.
The particle positions are also made periodic.
The overall size of the box roughly corresponds to the cutoff scale $b$ used in our analytic calculation (see Appendix~\ref{sec:analyticcalc}).

\subsection{Physical Scenario}

In the simulations presented below, we consider a massive object (either a point particle or Plummer sphere) traveling through an initially uniform ULDM background that may or may not have self-interactions.
Each simulation has the same initial velocity $v_{\rm{rel}}$ and background density, and we control the dimensionless parameter quantifying the expected dynamical friction force, $\xi$ (Eq.~\ref{xi}), by varying $\kappa$.  
We also ran simulations to verify the numerical convergence of the code; these results are summarized in Appendix~\ref{sec:restest}.

Simulations with large attractive self-interactions are not straightforward to simulate.  In particular, gravitational wakes will collapse if self-interactions are too large, which our simulations are not designed to study.  For this reason, we explore a wider range of repulsive than attractive self-interactions in this work.

The following sections contain measurements of the dynamical friction force due to a wake.
These forces are not constant---in particular, the particle starts in a homogeneous background in which it experiences no acceleration.
However, the wake immediately begins to grow, and the particle experiences a deceleration.
In the parameter ranges we examine, there is minimal positive feedback between the growth of the wake and the deceleration of the particle, and the deceleration is small enough during the length of the simulations such that the system approaches a steady state; see Appendix~\ref{sec:restest} for details.

For illustration, the left panel of Fig.~\ref{velvstime} shows how the velocity of a moving object will change as a function of time and the right panel of Fig.~\ref{velvstime} shows how the expected position of the point mass in a self-interacting background changes compared to when there are no self-interactions.  The solid lines in these plots show the region where the moving object is almost in steady state; in the following sections, we report the average acceleration in this nearly-steady state.

\section{Results}
\label{sec:results}

\subsection{Parameter Space Exploration}
\label{sec:theoryresults}

\begin{figure*}[t]
    \centering
    \includegraphics[width=.96\textwidth]{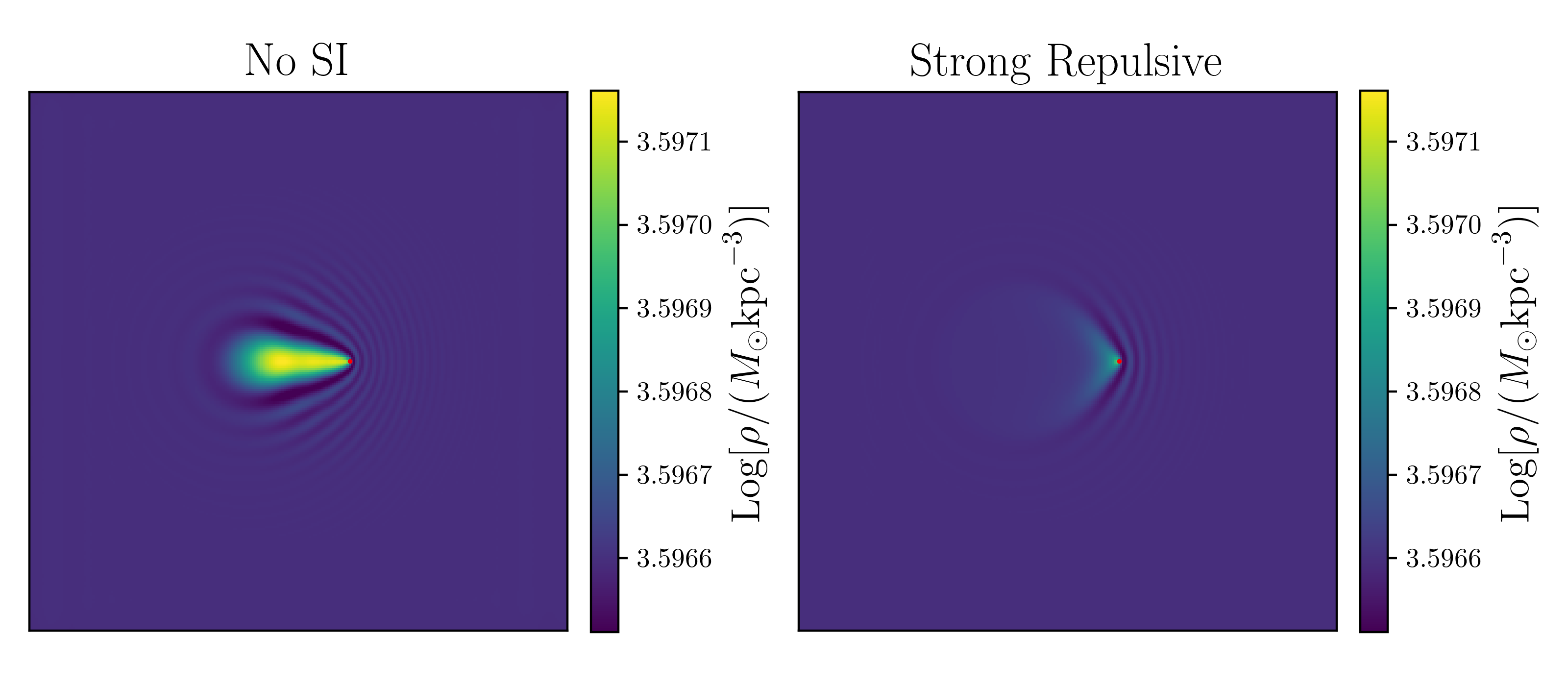}
    \caption{Snapshots of a point mass traveling through a uniform ULDM background with zero self-interactions (left panel; $\xi = 0$) and strong, repulsive self-interactions (right panel; $\xi = 2$).  These cases correspond to self-interaction strengths of $\kappa = 0$ and $\kappa = 1230$, respectively.  The red markers on the plots show the position of the point mass.  Gravitational wakes in the non-interacting case are denser than those in the presence of repulsive self-interactions.
    A corresponding animation can be found at \url{https://www.youtube.com/watch?v=T5Xq7muPq8A} and is permanently archived at \url{https://doi.org/10.5281/zenodo.7927475}.
    }
    \label{DFexample}
\end{figure*}

We first explore how the magnitude of the dynamical friction force varies as a function of ULDM self-interaction strength, and the scaling between these quantities.
In each of the following simulations, we use a mass of $4.4\times10^{5}~M_{\odot}$, a box length of 15 kpc, an initial relative velocity of $1.9\times10^{4}~\mathrm{km~s^{-1}}$, and a background density of $4.0\times10^{-6}~M_{\odot}~\mathrm{pc^{-3}}$.\footnote{The Jeans length for our fiducial cosmology and ULDM models is of $\mathcal{O}(100~\mathrm{kpc})$; our box is much smaller than this, and is therefore not unstable to gravitational collapse.}
When we use a Plummer sphere in this section, the Plummer radius is $a=0.0002 \;\mathcal{L}$.  In practice, these parameters (i.e., a large relative velocity, small object mass, and low background density) will make the measured dynamical friction force very small; we choose them to explore the phenomenological changes self-interactions could have on dynamical friction rather than to model particular astrophysical scenarios, as in the following sections.

Fig.~\ref{DFexample} shows snapshots of two simulations where a point particle travels through a uniform ULDM background with either zero or strong, repulsive self-interactions.  Here, the repulsive self-interaction strength of $\kappa=1230$ is chosen to highlight the potential impact of this effect, rather than to fall within the observational constraints described in Sec.~\ref{sec:constraints}. Note that these simulations span a fairly small dynamic range of densities because of the idealized parameter choices above; this does not affect the interpretation of our results.

\begin{figure}
    \centering
    \includegraphics[width=.48\textwidth]{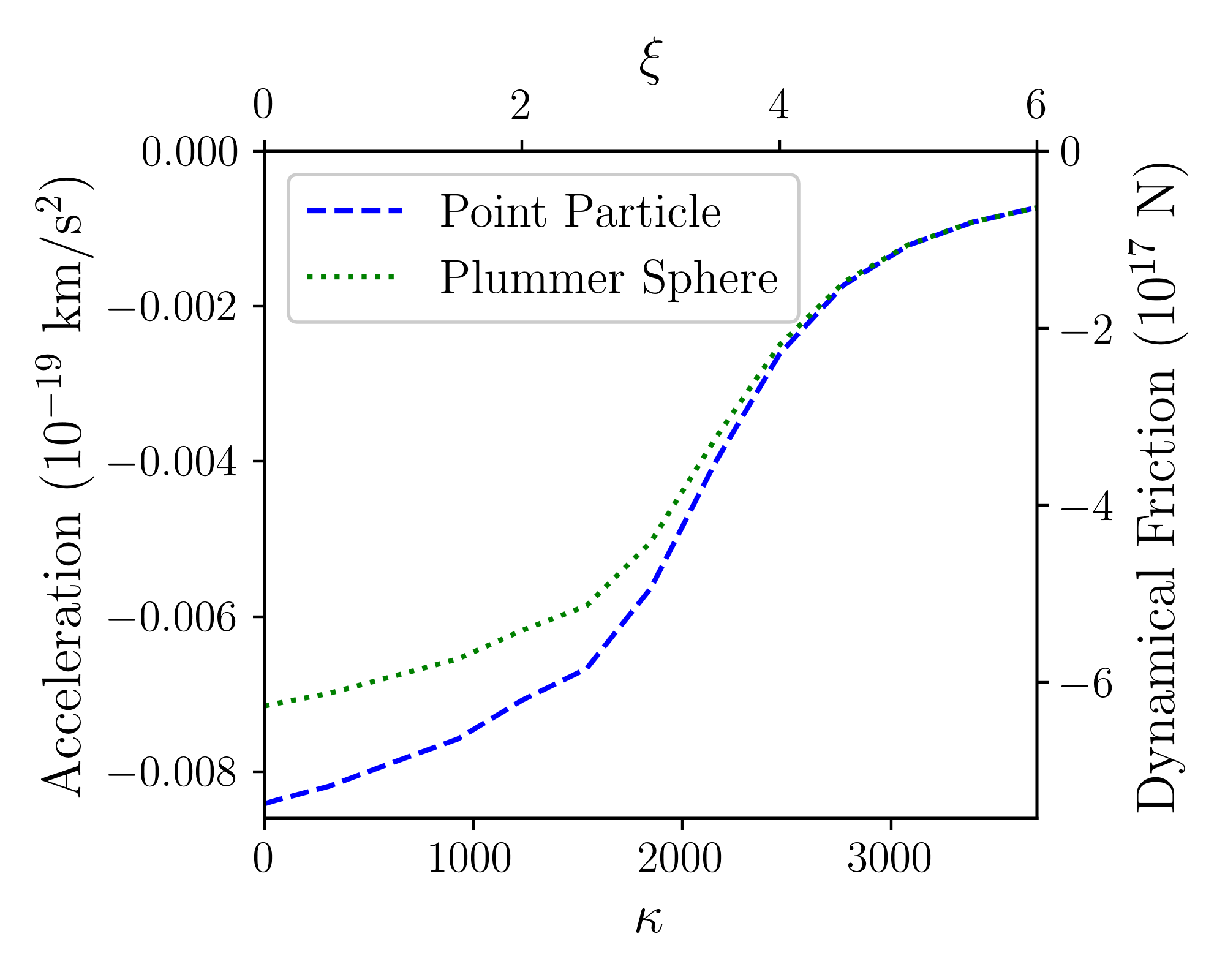}
    \caption{Dynamical friction force versus repulsive self-interaction strength $\kappa$ for a point mass (blue dashed) and Plummer sphere (green dotted) traveling through an initially uniform ULDM background.  The masses for the point mass and Plummer sphere are both $4.4\times10^{5}~M_{\odot}$.  The Plummer sphere experiences a smaller dynamical friction force for all self-interaction strengths, although the two cases converge for strong, repulsive self-interactions. The panels of Fig.~\ref{DFexample} correspond to the $\xi=0$ and $\xi=2$ points on the blue line, and the panels of Fig.~\ref{zoomin} correspond to the $\xi=0$ points on the blue and green lines.}
    \label{plummertest}
\end{figure}

We find that gravitational wakes are denser when the self-coupling is zero compared to the strong, repulsive case, in agreement with the intuition discussed in Sec.~\ref{sec:intro}, that repulsive self-interactions ``smooth out'' density wakes. Quantitatively, Fig.~\ref{plummertest} shows results from a suite of point-mass simulations as a function of the repulsive self-interaction strength. We identify three regimes: ($i$) for $0\lesssim \xi \lesssim 3$, the dynamical friction force slowly decreases as self-interactions become more repulsive (as $\xi\rightarrow 0$, our simulation prediction and analytic calculation agree; see Appendix~\ref{sec:analyticcalc}); ($ii$) for $3\lesssim \xi \lesssim 4$, the dynamical friction force decreases roughly linearly with the repulsive self-interaction strength; and ($iii$) for $\xi\gtrsim 4$, the dynamical friction force approaches zero, consistent with our analytic expectation.

\begin{figure*}
    \centering
    \includegraphics[width=.96\textwidth]{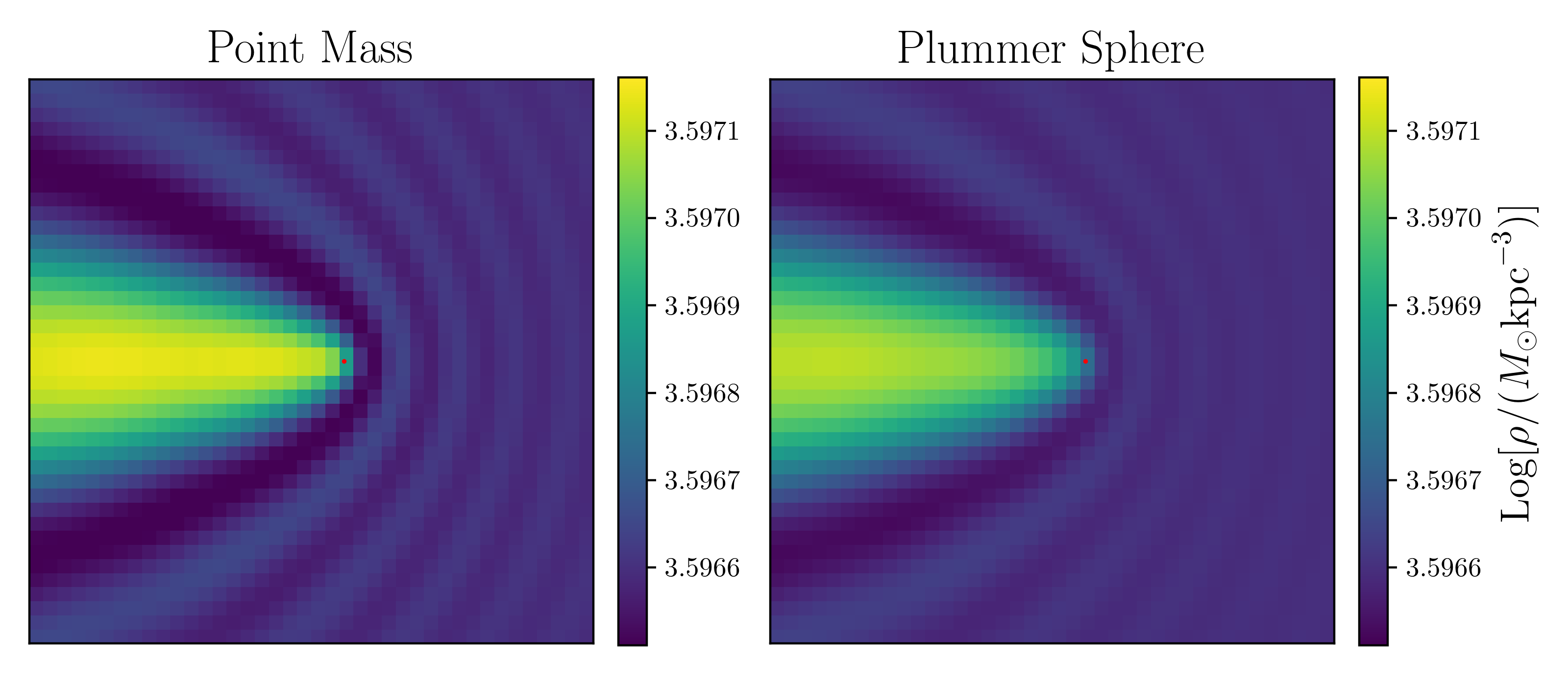}
    \caption{Zoomed-in snapshots of a point mass (left panel) and Plummer sphere (right panel) moving through a uniform ULDM background without self-interactions.  The red markers show the position of the point mass and Plummer sphere.  Each figure spans about 2 kpc.  Note that the left panel corresponds to a zoomed-in region from the left panel of Fig.~\ref{DFexample}.  The density contrast is slightly larger in the point mass case, consistent with our finding that the dynamical friction force is smaller for extended mass profiles like the Plummer sphere.  
    }
    \label{zoomin}
\end{figure*}

We also simulate Plummer sphere profiles to test whether an extended mass distribution responds differently to self-interactions than point masses.
Snapshots of these simulations look very similar to those in Fig.~\ref{DFexample}; as expected, variations in the dark matter density are slightly smoother compared to the point mass simulations.  A zoomed-in comparison between the point mass and Plummer sphere simulations in a case without self-interactions is shown in Fig.~\ref{zoomin}.

In Fig.~\ref{plummertest}, we compare the acceleration due to dynamical friction against $\xi$ for the point particle and Plummer sphere cases.
The comparison is reasonable, in that the Plummer sphere experiences a smaller dynamical friction force for all $\xi$ because it is less compact and thus generates a smaller wake.
However, this difference between the accelerations of point masses and Plummer spheres is much smaller for stronger repulsive self-interactions.
In the large $\xi$ limit, both accelerations tend towards zero, consistent with our analytic calculation. In particular, dynamical friction vanishes due to the sound speed---which is a property of the ULDM background, rather than the moving object---so this prediction is not sensitive to the profile assumed for the test mass. 

These results confirm many of our qualitative expectations for dynamical friction in the presence of ULDM self-interactions in a test scenario. We now turn to simulations with more realistic parameter choices in order to quantify these effects in various astrophysical settings.

\subsection{Dynamical Friction on Supermassive Black Holes}
\label{sec:realisticresults}

\begin{figure*}[t!]
    \centering
    \includegraphics[width=.48\textwidth]{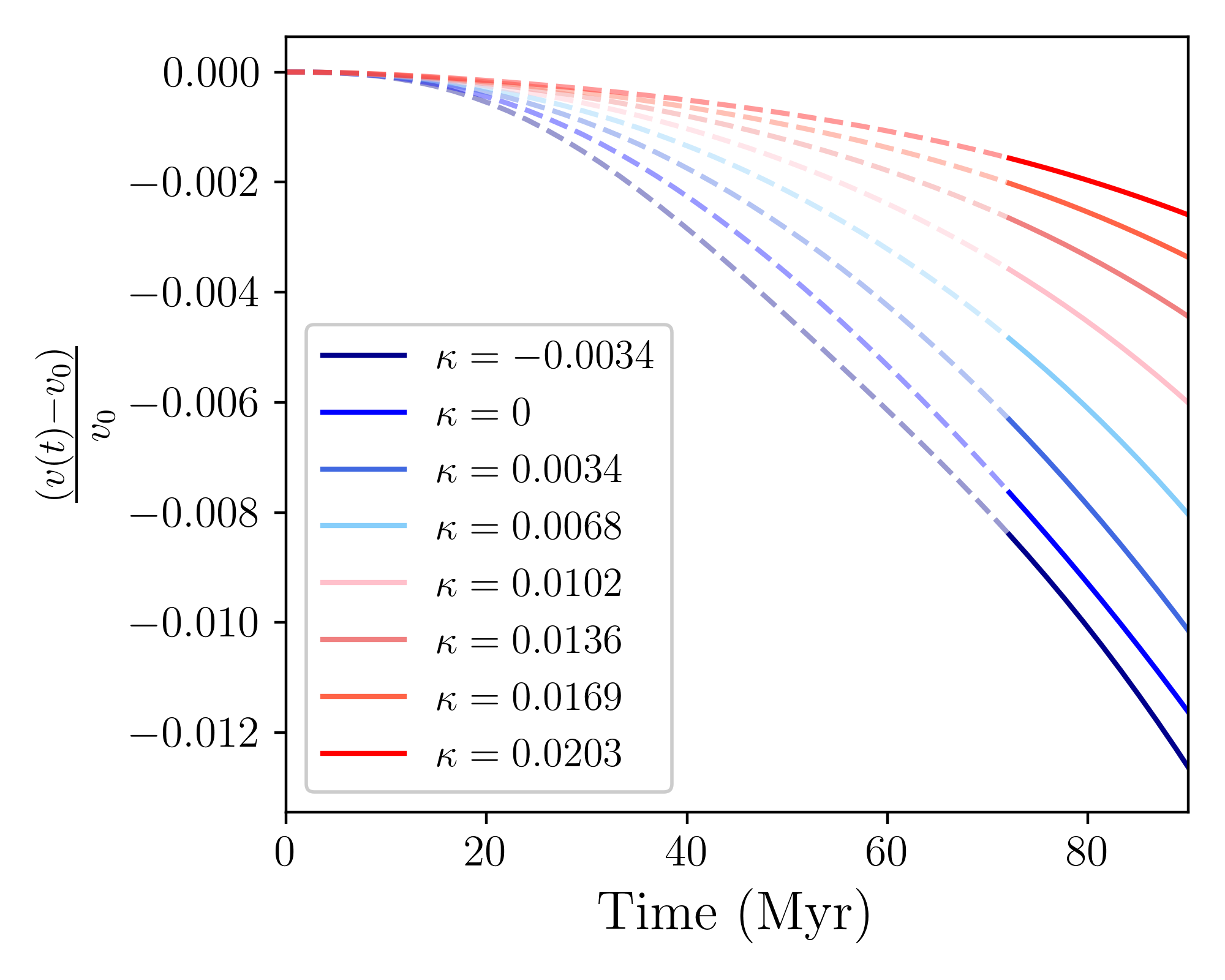}
    \includegraphics[width=.48\textwidth]{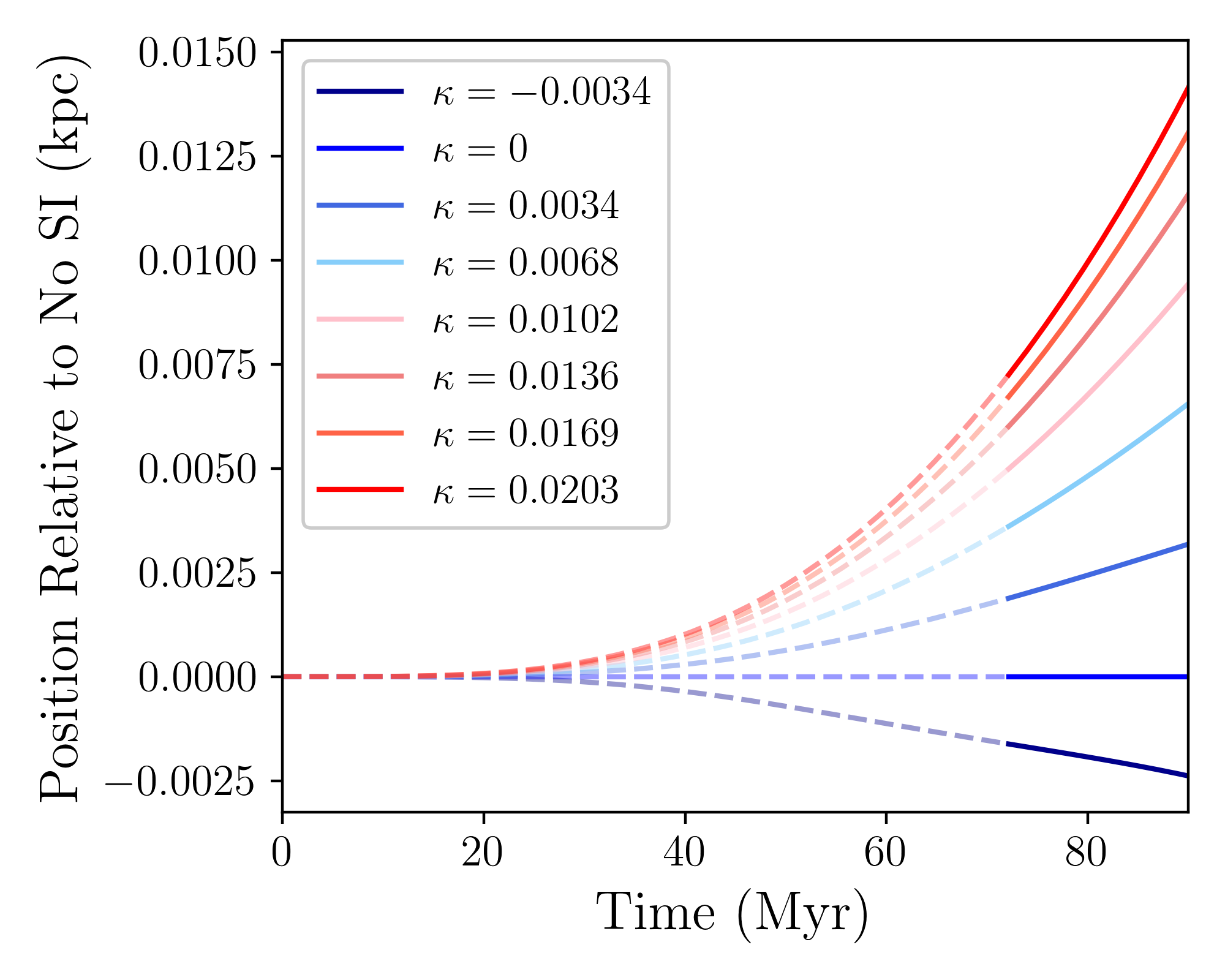}
    \caption{Left panel:  fractional change in velocity as a function time for a system representative of an SMBH traveling through a uniform ULDM background.  The solid lines represent the portion of the data where we assume the point mass is approximately in a steady state to calculate the deceleration due to dynamical friction. Note that these results are qualitatively similar to Fig.~\ref{velvstime}; however, quantitatively, the fractional change in velocity is much greater and it takes much longer to reach a steady state for the SMBH simulations due to its lower relative velocity and the higher background density. Right panel: the difference in the expected position, as a function of time, for an SMBH traveling through ULDM backgrounds with various self-interaction strengths, plotted relative to the position of the same SMBH in a non-interacting scenario. 
    }
    \label{velvstimebh}
\end{figure*}

\begin{figure*}
    \centering
    \includegraphics[width=.96\textwidth]{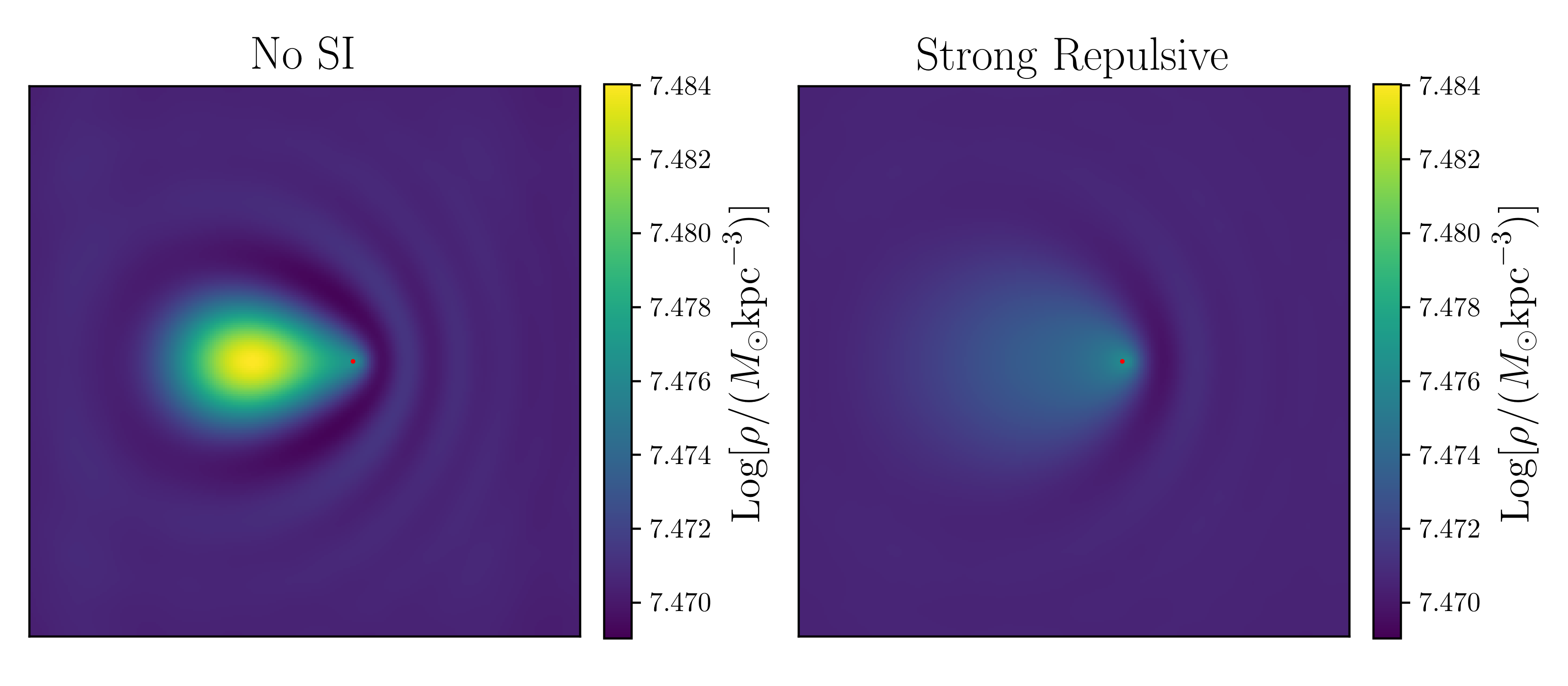}
    \caption{Snapshots of a Plummer sphere representing an SMBH traveling through a uniform ULDM background with either no self-interactions (left panel; $\xi = 0$) or strong, repulsive self-interactions (right panel; $\xi = 2$).  These correspond to self-interaction strengths of $\kappa = 0$ and $\kappa = 0.0068$, respectively.  Red markers show the position of the point mass.  The box length is about $12~\rm{kpc}$ across.  In this scenario, noticeable differences in the gravitational wakes arise when modeling values of $\kappa$ that are largely consistent with existing observational constraints.
        A corresponding animation can be found at \url{https://www.youtube.com/watch?v=27LzOpsgXxs} and is permanently archived at \url{https://doi.org/10.5281/zenodo.7927475}.
    }
    \label{DFbhexample}
\end{figure*}

\begin{figure}
    \centering
    \includegraphics[width=.48\textwidth]{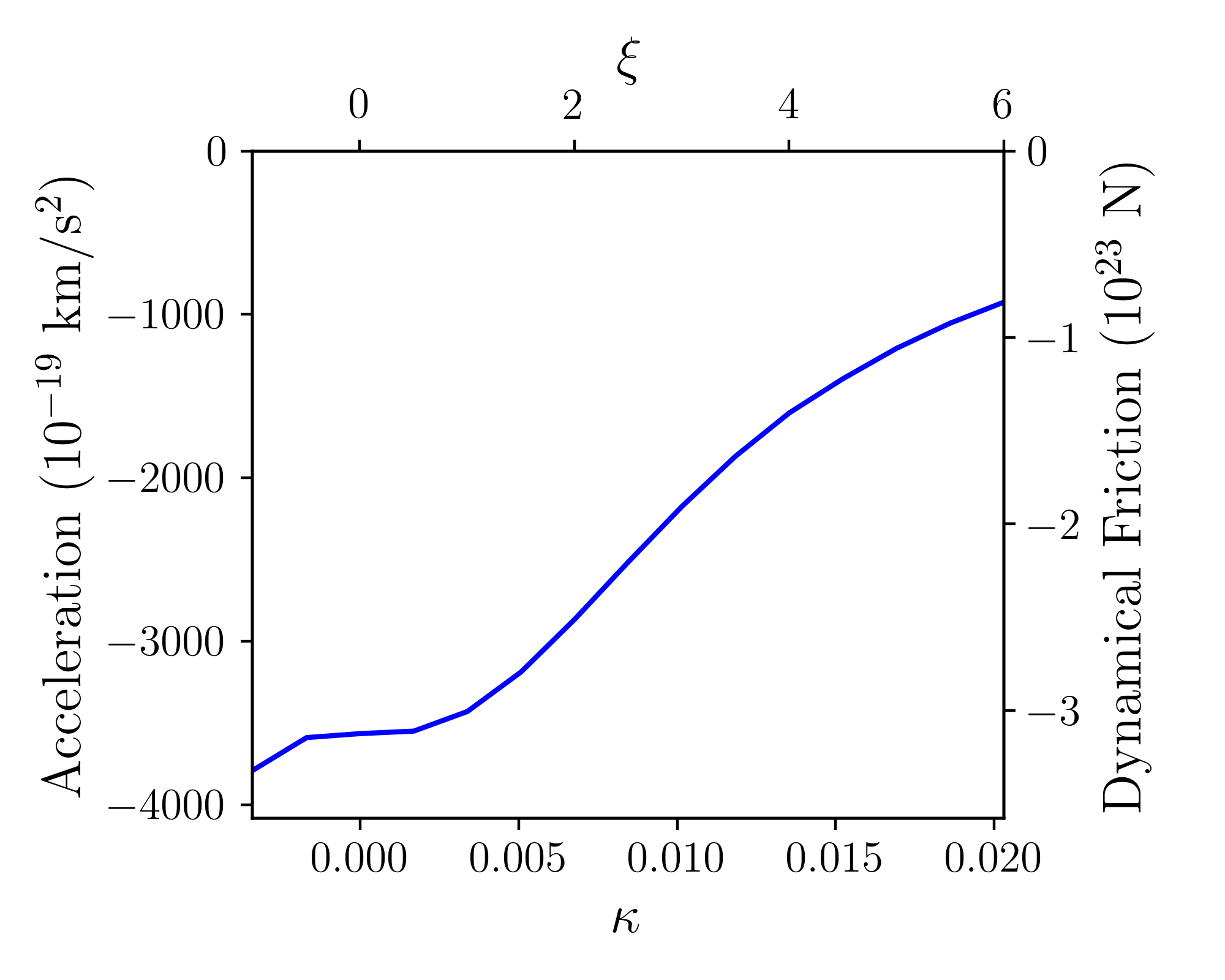}
    \caption{Acceleration versus self-interaction strength for a Plummer sphere (representing an SMBH) traveling through a uniform ULDM background.  This set of simulations uses a greater background density and smaller relative velocity compared to the simulations used in Fig.~\ref{plummertest}, which increases the dynamical friction force experienced by the Plummer sphere.  Note that, in this setup, the dynamical friction force approaches zero at a slower rate when $\xi$ increases compared to the simulations used to generate Fig.~\ref{plummertest}.
    }
    \label{bhtest}
\end{figure}

Ref.~\cite{Wang:2021udl} studied the dynamical friction experienced by an SMBH traveling through an ULDM background near the center of a soliton. We study the effects of ULDM self-interactions in a similar scenario, in which we initialize a Plummer sphere with a mass of $9.6\times10^{5}~M_{\odot}$ and Plummer radius of  $0.04~\rm{kpc}$  traveling with velocity $v = 50~\rm{km~s^{-1}}$ through a uniform background density of $\bar{\rho} = 0.0295~M_{\odot}~\rm{pc}^{-3}$; we use a box length of $12~\mathrm{kpc}$ for these simulations. These values are chosen to be representative of an SMBH near the center of a dark matter halo.  We plot velocity versus time in the left panel of Fig.~\ref{velvstimebh}, and we again indicate the region where we assume the SMBH is in a steady state.  The position of the SMBH for different self-interactions relative to the position of an SMBH in a non-interacting scenario is shown in the right panel of Fig.~\ref{velvstimebh}; the resulting accelerations  are plotted in Fig.~\ref{bhtest}, and two snapshots from these simulations are shown in Fig.~\ref{DFbhexample}.  When comparing Fig.~\ref{bhtest} to Fig.~\ref{plummertest}, we see that the dynamical friction force approaches zero more slowly in the SMBH scenario.  This follows because the dimensionless quantum mach number, $M_{\mathcal{Q}}$~\cite{Lancaster:2019mde}, is much smaller in our SMBH simulations relative to those in the previous section.  Our analytic calculation that predicts the dynamical friction force vanishes at $\xi=4$ is only valid in the limit where $M_{\mathcal{Q}}$ is large.  In Fig.~\ref{plummertest}, $M_{\mathcal{Q}} \approx 2 \times 10^5$, whereas in Fig.~\ref{bhtest}, $M_{\mathcal{Q}} \approx 200$.

The self-interactions for these SMBH tests span $-1<\xi<4$, and thus include both attractive and repulsive scenarios.  However, we note that the typical self-interaction magnitudes in these tests are much smaller than in the previous section due to the SMBH's relatively low velocity and the relatively high background density. When varying $\kappa$ through the observationally unconstrained interval $-0.0034 < \kappa < 0.0203$, we find that the acceleration due to dynamical friction varies by a factor of $\sim 4$. Thus, self-interactions in an ULDM background may significantly impact the acceleration experienced by an SMBH, and the time scale on which it sinks to the center of the halo.

According to Ref.~\cite{Wang:2021udl}, the orbital lifetime for an SMBH on a circular orbit around a solitonic core is roughly
\begin{equation}
    \tau = \frac{L}{r \abs{F_{\mathrm{DF}}}}
\end{equation}
where $L$ is the angular momentum of the orbit, $r$ is the radius, and $\abs{F_{\mathrm{DF}}}$ is the dynamical friction force.  Thus, typical repulsive self-interactions we study are expected to increase SMBH orbital lifetimes by factors of a few, compared to scenarios with no self-interactions. For sufficiently large repulsive self-interactions, the orbital lifetime may formally diverge as the dynamical friction force approaches zero, although we note that $\abs{F_{\mathrm{DF}}}$ vanishes more slowly in the SMBH case than for the test case in the previous section.

\subsection{Dynamical Friction on Massive Satellite Galaxies}

Satellite galaxies that occupy dark matter halos with sizable masses, compared to the mass of the host halo they fall into, experience significant dynamical friction in CDM. 
We therefore study how a system like the Large Magellanic Cloud (LMC), which is only a few times (up to $\sim 1$ order of magnitude) less massive than the Milky Way (MW) system it recently fell into~\cite{Erkal181208192,Shipp210713004}, might evolve differently in a self-interacting ULDM background. 
Recent work has compared dynamical friction on the LMC in non-self-interacting ULDM and in particle CDM, and found that although the wakes have difference structures, the dynamical friction forces are similar~\cite{Foote:2023exq}.

Ideally, we would model the LMC as a soliton moving through a self-interacting background or self-consistently accreting into a ULDM halo that represents the MW. Although \texttt{UltraDark.jl} is capable of simulating a soliton traveling through a uniform ULDM background, we leave a detailed study of this case to future work and instead extrapolate results from the simulations presented above to bracket the range of possible LMC behavior.

In the scenario without self-interactions, the acceleration due to dynamical friction for the LMC obtained by extrapolating our simulation results above is $\mathcal{O}(10^{-14}~\mathrm{km~s^{-2}})$.  This assumes a approximate background density of $\bar{\rho} = 2\times10^{-4}~M_{\odot}~\mathrm{pc}^{-3}$, a relative velocity of $v_{\mathrm{rel}} = 100~\mathrm{km~s}$, and an LMC mass of $10^{11}~M_{\odot}$.  The force predicted by Eq.~\ref{DFresult}, which assumes a point object, gives an acceleration of $4\times10^{-14}~\mathrm{km/s^2}$, in reasonable agreement with our simulation-based prediction.  We note that, because of the LMC's extended mass profile, the dynamical friction force predicted by Eq.~\ref{DFresult} is an overestimate. Furthermore, in this setup, $\xi=4$ (for which dynamical friction would approximately vanish) corresponds to $\lambda= 4\times10^{-88}$, which is generally larger than the existing observational constraints on $\lambda$ summarized in Sec.~\ref{sec:constraints}. 

Thus, plausible values of repulsive ULDM self-interaction strengths are expected to lengthen the LMC's orbital decay time scale, but not to erase the dynamical friction force on it entirely. Quantitatively, the expected distance traveled by the LMC, over the roughly 1 Gyr since it fell into the MW~\cite{Kallivayalil:2013xb}, is at most $\approx100~\mathrm{kpc}$ more in an ULDM background with repulsive self-interactions on the order of $\lambda\approx 10^{-88}$ than in a non-interacting scenario.  This upper bound assumes that the background density the LMC traveled though is a constant $\bar{\rho} = 2\times10^{-4}~M_{\odot}/\mathrm{pc}^3$ and that LMC is in steady state.  In reality, the LMC started in a less dense region, meaning the dynamical friction force was originally smaller.\footnote{However, the LMC also loses mass during its orbit; a self-consistent simulation that includes tidal stripping is needed to quantify the relative importance of this effect.}

Given recent, high-precision astrometric measurements of MW satellite proper motions~\cite{Patel200101746,Pace220505699}, shifts of $\mathcal{O}(10~\mathrm{kpc})$ in expected orbits may be measurable, thereby probing ULDM self-interactions. We leave this exciting possibility to future work, noting that complementary observables---e.g., the dark matter wake induced by the LMC~\cite{Garavito-Camargo:2019kxw}---may also be used to search for the dynamical effects of self-interacting ULDM.

\section{Conclusions}
\label{sec:conclusions}

We have presented numerical pseudospectral simulations to demonstrate the effects of quartic self-interactions on dynamical friction in ULDM models.

To explore the relevant parameter space, we first studied a point mass traveling through a uniform ULDM background and quantified how the dynamical friction force experienced by the particle depends on a dimensionless parameter, $\xi$, that we derived as a useful proxy for the expected dynamical friction force in self-interacting ULDM models. The dynamical friction force monotonically decreases as the self-interaction becomes more repulsive, approaching zero in some scenarios; we confirmed the latter result with an analytic calculation.
This is expected because repulsive self-interactions suppress the formation of gravitational wakes. It is remarkable that the dynamical friction force \emph{vanishes} in sufficiently repulsive self-interaction scenarios.

Our simulations indicate that plausible ULDM self-interaction strengths can significantly impact the dynamics of astrophysical objects moving through an ULDM background. 
For example, we have shown that the orbital lifetimes of both central SMBHs and massive satellite galaxies like the LMC are affected at the $\mathcal{O}(1)$ level in ULDM models with plausible self-interaction strengths, relative to the case with no self-interaction. 
Moreover, there are likely other astrophysical scenarios---beyond the SMBH and satellite galaxy cases we have focused on here---in which the effects of ULDM self-interactions on dynamical friction are relevant. 
Identifying and simulating these cases will help place constraints on the allowed range of self-interactions in ULDM models.

There are a few caveats to our results. First, we used external gravitational potentials when modeling the acceleration of the moving bodies. While this does not necessarily impact our point particle simulations, our simulations using an extended Plummer sphere profile could be affected by this choice because the test particle profile would likely be deformed over the course of the simulation. In general, our tests using an extended profile instead of a point mass show that, as expected, the force of dynamical friction is smaller for an extended object compared to a point object, provided that the objects have identical masses.

Next, our results may be impacted slightly by how our simulations were initialized and executed.  In particular, in Appendix~\ref{sec:restest}, we show that our measurements are converged as a function of grid size for strong, repulsive self-interactions and that they are sensitive to numerical resolution at the $\sim 10\%$ level for zero or attractive self-interactions. We note that the choice of box size and periodic boundary conditions can often affect the results of pseudospectral simulations; however, we deliberately use a large box size to mitigate this source of error.

In future work, we plan to run more realistic ULDM simulations focused on the astrophysical systems studied here: SMBHs at the center of dark matter halos (similar to, e.g., Ref.~\cite{Wang:2021udl}) and massive satellite galaxies. An important step in this direction will be to self-consistently model solitons traveling through an ULDM background, and, ideally, to treat the ``background'' itself as an ULDM halo (including its central soliton). Finally, we note that Ref.~\cite{Lancaster:2019mde} quantify how dynamical friction in a non-interacting ULDM background is altered by a non-zero velocity dispersion, which we have not modeled here.  This effect also deserves a dedicated analysis in the presence of self-interactions.

\acknowledgments
We thank Arka Banerjee for his foundational role in this collaboration, and we thank JiJi Fan, Anthony Mirasola and Sebastian Wagner-Carena for helpful discussions.

This research made use of computational resources at SLAC National Accelerator Laboratory, a U.S.\ Department of Energy Office; the authors are thankful for the support of the SLAC computing team and all administrative and custodial staff at our respective institutions, including Katie Makem-Boucher and Michelle Mancini. This research was supported in part by the National Science Foundation under Grants No.\ NSF PHY-1748958 and No. 1929080.
This work was performed in part at Aspen Center for Physics, which is supported by National Science Foundation under Grant No. PHY-1607611. 
CPW thanks the late Karsten Pohl for actively supporting the application for NSF grant No.\ 1929080.

\appendix

\bibliography{masterlibrary}

\appendix

\section{Analytic Calculation} 
\label{sec:analyticcalc}

Continuing our calculation of the expected dynamical friction force in a self-interacting ULDM background from Eq.~\ref{xi}, we perform an inverse Fourier transform, which gives
\begin{equation}
    \alpha(\tilde{r}) = \frac{16\pi}{M_{\mathrm{Q}}} \int \frac{d^3 k}{(2\pi)^3} \frac{e^{i \tilde{k} \cdot \tilde{r}}}{(\tilde{k}^4 - 4\tilde{k}^2_z+\xi \tilde{k}^2)}.
\end{equation}
This leads to
\begin{equation}
    \begin{split}
        \alpha(\tilde{r}) = 
        & \frac{16\pi}{(2\pi)^2 M_{\mathrm{Q}}} \int^{\infty}_0 d\tilde{k}_R  \tilde{k}_R J_0(\tilde{k}_R \tilde{R}) 
        \\& \times \int^\infty_{-\infty} d\tilde{k}_z \frac{e^{i \tilde{k}_z \cdot \tilde{z}}}{(\tilde{k}^4 - 4\tilde{k}^2_z+\xi \tilde{k}^2)}.
    \end{split}
\end{equation}
We can evaluate the last part of the integral by writing
\begin{equation}
    \tilde{k}^4 - 4\tilde{k}^2_z+\xi \tilde{k}^2 = \tilde{k}^4_z + m \tilde{k}^2_z + n
\end{equation}
where $m = 2\tilde{k}_R^2+(\xi - 4)$ and $n = \tilde{k}^4_R + \xi \tilde{k}_R^2$.  We have also used $\tilde{k}^2 = \tilde{k}_R^2 + \tilde{k}_z^2$.  By defining
\begin{equation}
    \chi_\pm = \sqrt{\frac{-m\pm\sqrt{m^2-4n}}{2}}
\end{equation}
we can write the integral as
\begin{equation}
    \begin{split}
        I & = \int^\infty_{-\infty} d\tilde{k}_z \frac{e^{i \tilde{k}_z \cdot \tilde{z}}}{(\tilde{k}^4 - 4\tilde{k}^2_z+\xi \tilde{k}^2)} 
        \\ & = \int^\infty_{-\infty} d\tilde{k}_z \frac{e^{i \tilde{k}_z \cdot \tilde{z}}}{(\tilde{k}_z-\chi_-)(\tilde{k}_z + \chi_-)(\tilde{k}_z-\chi_+)(\tilde{k}_z+\chi_+)}.
    \end{split}
\end{equation}
There are four cases we need to consider to evaluate the integral.  These are when $\xi > 4$, $4>\xi>0$, $0>\xi>-4$, and $-4>\xi$.  

\subsection{Case 1 ($\xi>4$):}

\begin{figure}
    \centering
    \includegraphics[width=.48\textwidth,trim={3cm 4cm 3cm 4cm},clip]{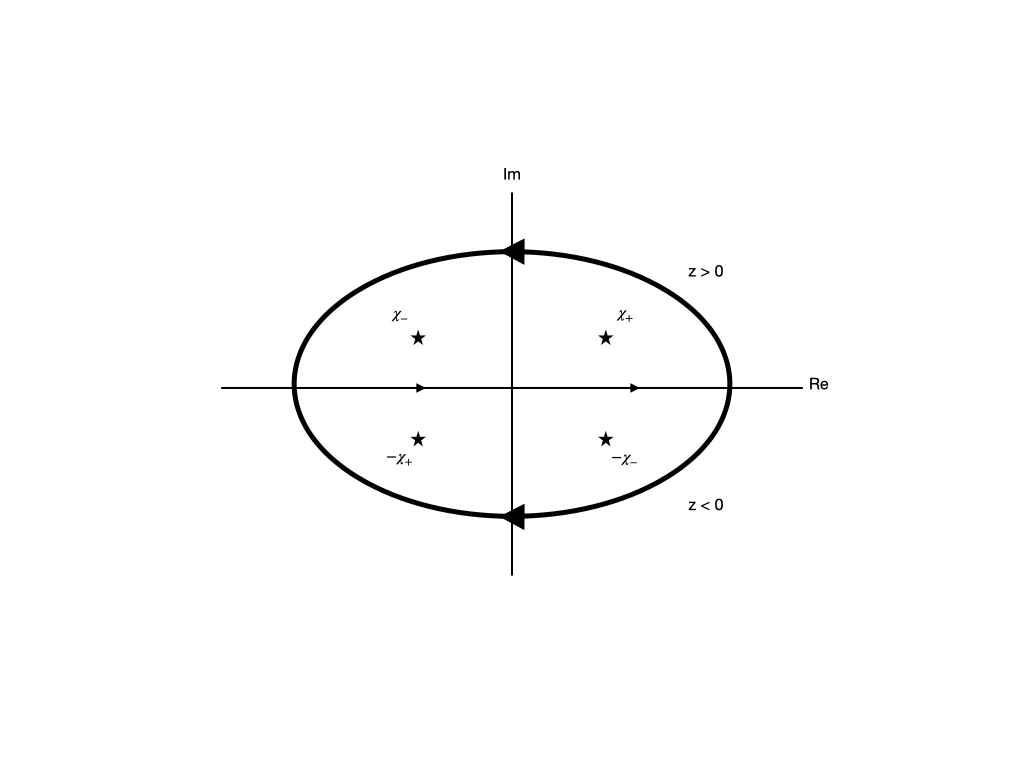}
    \caption[Case 1 (all pole off the real axis)]{Illustration of contours used for Case 1.}
    \label{Case1}
\end{figure}

In this first case, the pole always exists off the real axis.  There are two contours that need to be integrated.  A schematic of the contours used for this is in Fig.~\ref{Case1}. The first is the $z>0$ contour. We find
\begin{equation}
    I_1(\tilde{z}>0) = 2 \pi i \left(-\frac{\chi_-e^{i\chi_+ \tilde{z}} - \chi_+ e^{-i \chi_- \tilde{z}}}{2\chi_-^3 \chi_+ - 2 \chi_- \chi_+^3}\right).
\end{equation}
For $z<0$, we find
\begin{equation}
    I_1(\tilde{z}<0) = -2 \pi i \left(-\frac{\chi_-e^{-i\chi_+ \tilde{z}} - \chi_+ e^{i \chi_- \tilde{z}}}{2\chi_-^3 \chi_+ - 2 \chi_- \chi_+^3}\right).
\end{equation}
This can be summarized as
\begin{equation}
    I_1(\tilde{z}) = 2 \pi i \left(\frac{\chi_+e^{-i\chi_- \abs{\tilde{z}}} - \chi_- e^{i \chi_+ \abs{\tilde{z}}}}{2\chi_-^3 \chi_+ - 2 \chi_- \chi_+^3}\right).
\end{equation}
Since this expression is even in $z$, the contribution to the dynamical friction force is zero.

\subsection{Case 2 ($4>\xi>0$):}

There are two sub-cases that need to be considered for this case.  The first is when both $\chi_+$ and $\chi_-$ are real which happens when $0<\tilde{k}_R<1-\frac{\xi}{4}$.  The second sub-case is when $\chi_+$ and $\chi_-$ are both imaginary which happens if $\tilde{k}_R>1-\frac{\xi}{4}$.

\subsubsection{Case 2(a) ($0<\tilde{k}_R<1-\frac{\xi}{4}$):}

\begin{figure}
    \centering
    \includegraphics[width=.48\textwidth,trim={3cm 4cm 3cm 4cm},clip]{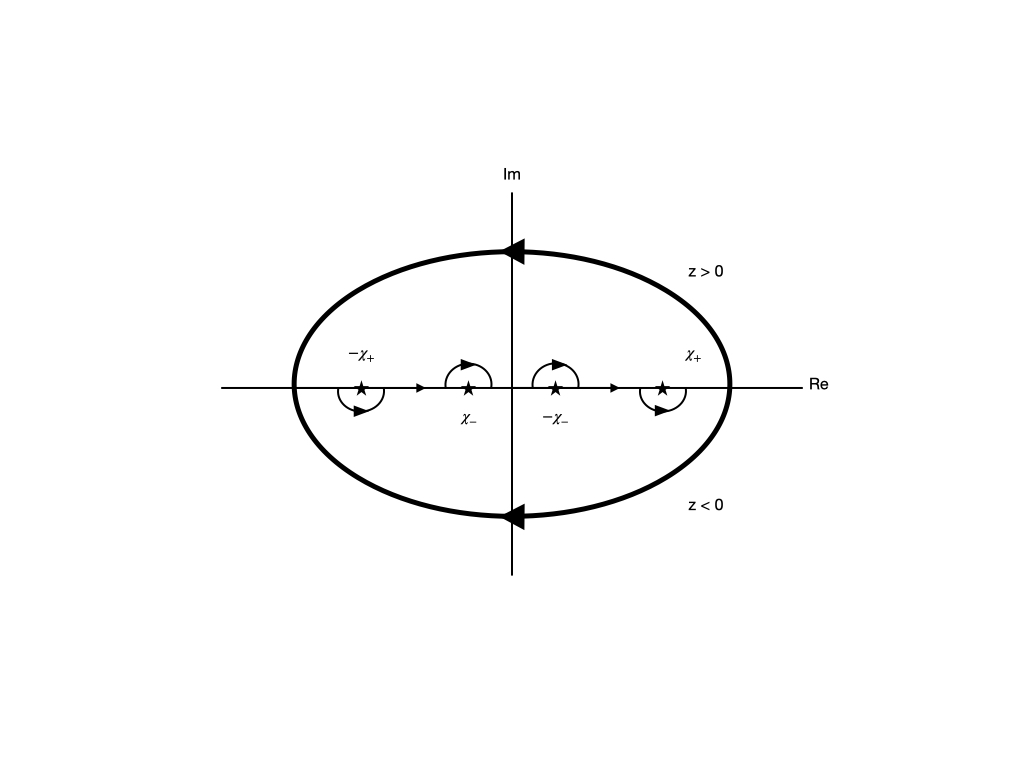}
    \caption[Case 2(a) (all poles on the real axis)]{Illustration of contours used for Case 2(a).}
    \label{Case2a}
\end{figure}

In this sub-case, there again are two contours we need to integrate over.  A schematic of this case is shown in Fig.~\ref{Case2a}.  We have
\begin{equation}
    I_{2a}(\tilde{z}>0) = 2\pi i \frac{i \sin(\chi_+ \tilde{z})}{\chi_+(\chi_+^2-\chi_-^2)}
\end{equation}
and
\begin{equation}
    I_{2a}(\tilde{z}<0) = - 2\pi i \frac{i \sin(\chi_- \tilde{z})}{\chi_-(\chi_-^2-\chi_+^2)}
\end{equation}
Putting this back into Eq. 14 gives
\begin{equation}
    \begin{split}
        \alpha_{2a} & = -\frac{16\pi}{(2\pi)^2M_{\mathcal{Q}}}\int_0^{1-\frac{\xi}{4}}d\tilde{k}_R \tilde{k}_R J_0(\tilde{k}_R \tilde{R})
        \\ & \times (I_{2a}(\tilde{z}>0)+I_{2a}(\tilde{z}<0))
        \\ & = -\frac{8}{M_{\mathcal{Q}}}\Biggl[\int^{1-\frac{\xi}{4}}_0 d\tilde{k}_R \tilde{k}_R J_0(\tilde{k}_R \tilde{R}) \frac{\sin(\chi_+ \tilde{z})}{\chi_+(2\chi_+^2+m)} 
        \\ & + \int^{1-\frac{\xi}{4}}_0 d\tilde{k}_R \tilde{k}_R J_0(\tilde{k}_R \tilde{R}) \frac{\sin(\chi_- \tilde{z})}{\chi_-(2\chi_-^2+m)} \Biggr].
    \end{split}
\end{equation}
We next change the limits of integration.
$\tilde{k}_{R} = 0 \rightarrow \chi_+ = \sqrt{4-\xi},~\chi_- = 0$ and $\tilde{k}_{R} = 1-\frac{\xi}{4} \rightarrow \chi_+ = \sqrt{1-\frac{\xi^2}{16}},~\chi_- = \sqrt{1-\frac{\xi^2}{16}}$.  We also write
\begin{equation}
    \tilde{k}_R = \sqrt{\frac{-2\chi_{\pm}-\xi+\sqrt{16\chi_{\pm}^2+\xi^2}}{2}}
\end{equation}
and
\begin{equation}
    d \tilde{k}_R = \mp d\chi_{\pm}\frac{\sqrt{2}\chi_{\pm}\left(\frac{4}{\sqrt{\xi^2+16\chi_{\pm}^2}}-1\right)}{\sqrt{\sqrt{\xi^2+16\chi_{\pm}^2}-(\xi+2\chi_{\pm}^2)}}.
\end{equation}
Multiplying $d \tilde{k}_R$ by $\tilde{k}_R$ simplifies a bit to
\begin{equation}
    d \tilde{k}_R \tilde{k}_R = d \chi_{\pm} \chi_{\pm} \left(\frac{4}{\sqrt{16\chi^2_{\pm}+\xi^2}}-1\right).
\end{equation}
We now have
\begin{equation}
    \begin{split}
        \alpha_{2a} & = -\frac{8}{M_{\mathcal{Q}}}\Biggl[\int^{\sqrt{4-\xi}}_{\sqrt{1-\frac{\xi^2}{16}}} d\chi_+ J_0(\beta(\xi, \chi_+)\tilde{R})\frac{\sin(\chi_+ \tilde{z})}{4\chi_+} \\ & + \int^{\sqrt{1-\frac{\xi^2}{16}}}_0 d\chi_- J_0(\beta(\xi, \chi_-)\tilde{R})\frac{\sin(\chi_- \tilde{z})}{4\chi_-} \Biggr].
    \end{split}
\end{equation}
Here I have defined $\beta(\xi, x)$ as 
\begin{equation}
     \beta(\xi, x) = \sqrt{\frac{-2x-\xi+\sqrt{16x^2+\xi^2}}{2}}
\end{equation}By changing $\chi_+,~\chi_-\rightarrow x$, this becomes
\begin{equation}
    \alpha_{2a} = -\frac{2}{M_{\mathcal{Q}}}\int^{\sqrt{4-\xi}}_{0} dx J_0(\beta(\xi, x)\tilde{R})\frac{\sin(x \tilde{z})}{x}.
\end{equation}

\subsubsection{Case 2(b) ($\tilde{k}_R>1-\frac{\xi}{4}$):}

In this case, both $\chi_+$ and $\chi_-$ are imaginary and we have the same result as case 1.  This means $I_{2b} = 0$ and thus $\alpha_{2b} = 0$.

To summarize,
\begin{equation}
    \alpha_2 = \alpha_{2a} + \alpha_{2b} = -\frac{2}{M_{\mathcal{Q}}}\int^{\sqrt{4-\xi}}_{0} dx J_0(\beta(\xi, x)\tilde{R})\frac{\sin(x \tilde{z})}{x}.
\end{equation}
To find the dynamical friction force, we use
\begin{equation}
    F_{\mathrm{DF}} = \bar{\rho} \int d^3 x \alpha(x) (\hat{x}_{||}\cdot \nabla)U(x).
\end{equation}

In this case,
\begin{equation}
    \begin{split}
        & F_{\mathrm{DF}}(4>\xi>0) = 
        \\ & \bar{\rho} \int d^3 x \alpha_2(x) (\hat{x}_{||}\cdot \nabla)\left[-\frac{G M}{r}+ \frac{\hbar^3\lambda}{2m^4}\bar{\rho}\alpha_2\right].
    \end{split}
\end{equation}
\begin{equation}
    \begin{split}
        & F_{\mathrm{DF}}(4>\xi>0) = 
        \\ & 2\pi G M \bar{\rho} \lambdabar \int^{\infty}_0 d\tilde{R} \int^{\infty}_{-\infty} d\tilde{z}\frac{\tilde{R}\tilde{z}}{(\tilde{R}^2+\tilde{z}^2)^{3/2}}\alpha_2(\tilde{R}, \tilde{z})
        \\ &
        +\frac{4\pi\hbar^3\lambda\bar{\rho}\lambdabar}{M_{\mathcal{Q}}^2m^4}\int^{\infty}_0 d\tilde{R} \int^{\infty}_{-\infty}d\tilde{z}
        \\ & \times \left[\int^{\sqrt{4-\xi}}_0 dx J_0(\beta(\xi, x)\tilde{R}) \frac{\sin(x \tilde{z})}{x}\right]
        \\ &
        \times\left[\int^{\sqrt{4-\xi}}_0 dx J_0(\beta(\xi, x)\tilde{R})\cos{(x \tilde{z})}\right]
    \end{split}
\end{equation}
By defining
\begin{equation}
    F_{\mathrm{rel}} = 4\pi\bar{\rho}\left(\frac{G M}{v_{\mathrm{rel}}}\right)^2,
\end{equation}
we obtain
\begin{equation}
    \begin{split}
        & F_{\mathrm{DF}}(4>\xi>0) = 
        \\ & -F_{\mathrm{rel}} \int^{\infty}_0 d\tilde{R} \int^{\infty}_{-\infty} d\tilde{z}\frac{\tilde{R}\tilde{z}}{(\tilde{R}^2+\tilde{z}^2)^{3/2}}J_0(\beta(\xi, x)\tilde{R}) \frac{\sin(x \tilde{z})}{x}
        \\&
        + \frac{F_{\mathrm{rel}}}{2}\xi\int^{\infty}_0 d\tilde{R} \int^{\infty}_{-\infty}d\tilde{z}
        \\ & \times \left[\int^{\sqrt{4-\xi}}_0 dx J_0(\beta(\xi, x)\tilde{R}) \frac{\sin(x \tilde{z})}{x}\right]
        \\&
        \times\left[\int^{\sqrt{4-\xi}}_0 dx J_0(\beta(\xi, x)\tilde{R})\cos{(x \tilde{z})}\right]
    \end{split}
\end{equation}

\subsection{Case 3 ($0>\xi>-4$):}

There are three sub-cases that need to be considered for this case.  The first is when both $\chi_+$ and $\chi_-$ are real which happens when $\sqrt{-\xi}<\tilde{k}_R<1-\frac{\xi}{4}$.  The second sub-case is when $\chi_+$ are real and $\chi_-$ are imaginary which happens if $0<\tilde{k}_R<\sqrt{-\xi}$.  The last case is when both $\chi_+$ and $\chi_-$ are imaginary which happens when $\tilde{k}_R>1-\frac{\xi}{4}$.

\subsubsection{Case 3(a) ($\sqrt{-\xi}<\tilde{k}_R<1-\frac{\xi}{4}$):}

This sub-case is very similar to sub-case 2(a), however, the limits on $\tilde{k}_R$ are different.  Doing the same contour integration as in sub-case 2(a), we obtain
\begin{equation}
    \begin{split}
        \alpha_{3a} = & -\frac{8}{M_{\mathcal{Q}}}\Biggl[\int^{1-\frac{\xi}{4}}_{\sqrt{-\xi}} d\tilde{k}_R \tilde{k}_R J_0(\tilde{k}_R \tilde{R}) \frac{\sin(\chi_+ \tilde{z})}{\chi_+(2\chi_+^2+m)} 
        \\ & + \int^{1-\frac{\xi}{4}}_{\sqrt{-\xi}} d\tilde{k}_R \tilde{k}_R J_0(\tilde{k}_R \tilde{R}) \frac{\sin(\chi_- \tilde{z})}{\chi_-(2\chi_-^2+m)} \Biggr]
    \end{split}
\end{equation}
Changing the limits of integration, we do
$\tilde{k}_{R} = \sqrt{-\xi} \rightarrow \chi_+ = \sqrt{\xi+4},~\chi_- = 0$ and $\tilde{k}_{R} = 1-\frac{\xi}{4} \rightarrow \chi_+ = \sqrt{1-\frac{\xi^2}{16}},~\chi_- = \sqrt{1-\frac{\xi^2}{16}}$.

By doing the same steps as in earlier cases, we arrive at
\begin{equation}
    \alpha_{3a} = -\frac{2}{M_{\mathcal{Q}}}\int^{\sqrt{\xi+4}}_{0} dx J_0(\beta(\xi, x)\tilde{R})\frac{\sin(x \tilde{z})}{x}.
\end{equation}

\subsubsection{Case 3(b) ($0<\tilde{k}_R<\sqrt{-\xi}$):}

\begin{figure}
    \centering
    \includegraphics[width=.48\textwidth,trim={3cm 4cm 3cm 4cm},clip]{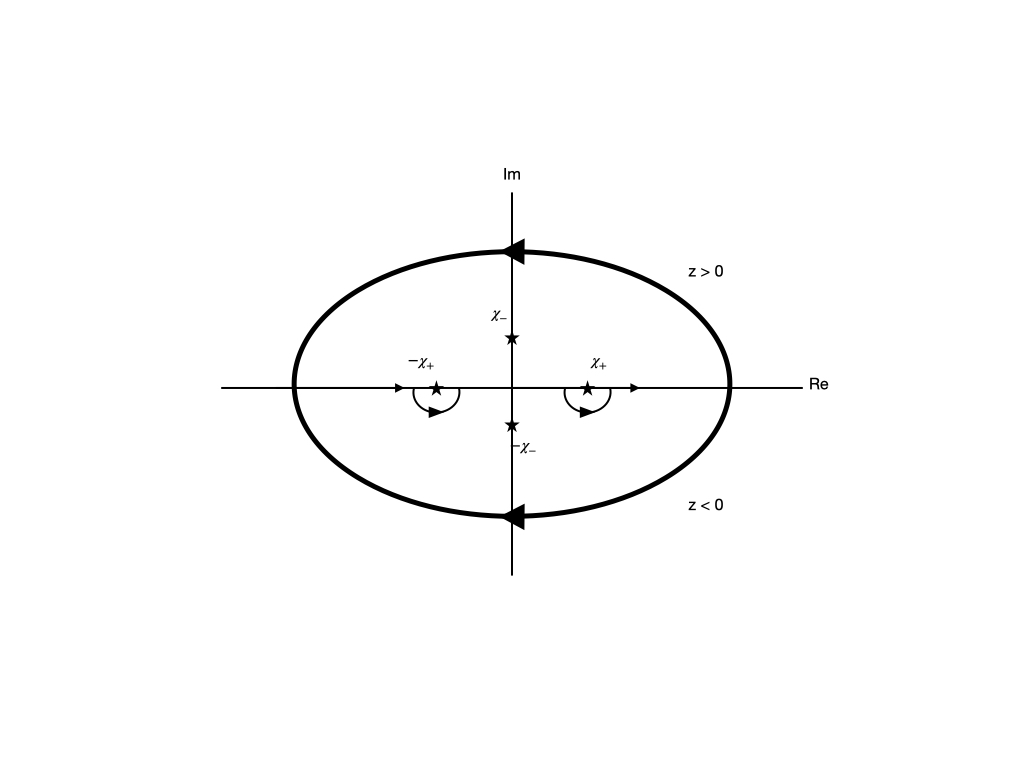}
    \caption[Case 3(b) (only $\chi_+$ exist on the real axis)]{Illustration of contours used for Case 3(b).}
    \label{Case3b}
\end{figure}

The schematic for this integral is given in Fig.~\ref{Case3b}.  In this sub-case, the two contour integrals give
\begin{equation}
    \begin{split}
        I_{3b}(\tilde{z}>0) & = 2\pi i \left[\frac{\chi_+ e^{i\chi_- \tilde{z}}-2i\chi_-\sin(\chi_+\tilde{z})}{2\chi_-^3\chi_+-2\chi_-\chi_+^3}\right]
        \\ & =\frac{\pi i \chi_+ e^{i \chi_- \tilde{z}}}{\chi_-\chi_+(\chi_-^2-\chi_+^2)} + \frac{2\pi\sin(\chi_+\tilde{z})}{\chi_+(\chi_-^2-\chi_+^2)}
    \end{split}
\end{equation}
and
\begin{equation}
    \begin{split}  
        I_{3b}(\tilde{z}<0) & = -2\pi i \left[-\frac{\chi_+ e^{-i \chi_-\tilde{z}}}{2\chi_-^3\chi_+-2\chi_-\chi_+^3}\right]
        \\ & = \frac{\pi i \chi_+ e^{-i \chi_- \tilde{z}}}{\chi_-\chi_+(\chi_-^2-\chi_+^2)}.
    \end{split}
\end{equation}
We can write $I_{3b}$ as
\begin{equation}
    I_{3b} = I_{3b,~\mathrm{even}} + I_{3b,~\mathrm{odd}}.
\end{equation}
Again, the contribution from the even part is zero meaning that
\begin{equation}
    I_{3b} = -\frac{2\pi\sin(\chi_+ \tilde{z})}{\chi_+(\chi_+^2-\chi_-^2)}.
\end{equation}
After some manipulation, we then find
\begin{equation}
    \alpha_{3b} = \frac{8}{M_{\mathcal{Q}}}\int^{\sqrt{-\xi}}_{0} dx J_0(\tilde{k}_R\tilde{R})\frac{\sin(\chi_+ \tilde{z})}{\chi_+(2\chi_+^2+m)}.
\end{equation}
We need to change the integration limits again.  $\tilde{k}_{R} = 0 \rightarrow \chi_+ = \sqrt{4-\xi}$.  $\tilde{k}_{R} = \sqrt{-\xi} \rightarrow \chi_+ = \sqrt{\frac{\xi+4+\abs{\xi+4}}{2}}$.  Note that, in this case, if $\xi>-4$, $\chi_+ = \sqrt{\xi+4}$, otherwise $\chi_+=0$.  This results in
\begin{equation}
    \alpha_{3b} = -\frac{2}{M_{\mathcal{Q}}}\int^{\sqrt{4-\xi}}_{\sqrt{4+\xi}} dx J_0(\beta(\xi, x)\tilde{R})\frac{\sin(x \tilde{z})}{x}.
\end{equation}

\subsubsection{Case 3(c) ($\tilde{k}_R>1-\frac{\xi}{4}$):}

In this sub-case, both $\chi_+$ and $\chi_-$ are imaginary so the integrals are identical to case 1.  The integral is even in $\tilde{z}$ and does not contribute.

In total, we have
\begin{equation}
    \begin{split}
        \alpha_{3} & = \alpha_{3a}+\alpha_{3b}+\alpha_{3c} 
        \\& 
        = -\frac{2}{M_{\mathcal{Q}}}\int^{\sqrt{4+\xi}}_{0} dx J_0(\beta(\xi, x)\tilde{R})\frac{\sin(x \tilde{z})}{x} 
        \\ & +  -\frac{2}{M_{\mathcal{Q}}}\int^{\sqrt{4-\xi}}_{\sqrt{4+\xi}} dx J_0(\beta(\xi, x)\tilde{R})\frac{\sin(x \tilde{z})}{x}
    \end{split}
\end{equation}
or
\begin{equation}
    \alpha_{3} = -\frac{2}{M_{\mathcal{Q}}}\int^{\sqrt{4-\xi}}_{0} dx J_0(\beta(\xi, x)\tilde{R})\frac{\sin(x \tilde{z})}{x}
\end{equation}
which is the same expression for $\alpha_2$.  Thus,
\begin{equation}
    F_{\mathrm{DF}}(0>\xi>-4) = F_{\mathrm{DF}}(4>\xi>0)
\end{equation}

\subsection{Case 4 ($\xi<-4$):}

There are two sub-cases that need to be looked at.  The first is when $\chi_+$ are real and $\chi_-$ are imaginary which happens when $0<\tilde{k}_R<\sqrt{-\xi}$.  The second sub-case is when $\chi_+$ and $\chi_-$ are imaginary which happens if $\tilde{k}_R>\sqrt{-\xi}$.
    
\subsubsection{Case 4(a) ($0<\tilde{k}_R<\sqrt{-\xi}$):}

This sub-case is identical to sub-case 3(b).  This gives $I_{4a}=I_{3b}$.

\subsubsection{Case 4(b) ($\tilde{k}_R>\sqrt{-\xi}$):}

Since both $\chi_+$ and $\chi_-$ are imaginary, for the same reason as previous cases, the contribution is zero.

In summarizing this case, we find $\alpha_2 = \alpha_3 = \alpha_4$ and
\begin{equation}
    F_{\mathrm{DF}}(-4>\xi) = F_{\mathrm{DF}}(0>\xi>-4) = F_{\mathrm{DF}}(4>\xi>0) 
\end{equation}

\subsection{Result}

The dynamical friction force is then
\begin{equation}
F_{\text{DF}}(\xi,\tilde{b}) = 
\left\{
    \begin{array}{lr}
        F_1(\xi,\tilde{b})+F_2(\xi,\tilde{b}), & \text{if } \xi < 4\\
       0, & \text{if } \xi > 4
    \end{array}
\right\}
\label{DFresult}
\end{equation}
where
\begin{equation}
    \begin{split}
        F_1(\xi, \tilde{b}) = 
        & -F_{\text{rel}}\int^{\tilde{b}}_0 d\tilde{R}\int^{\sqrt{\tilde{b}^2-\tilde{R}^2}}_{0}d\tilde{z}
        \\& \times \int^{\sqrt{4-\xi}}_0 dx \frac{\tilde{R}\tilde{z}}{(\tilde{R}^2+\tilde{z}^2)^{3/2}}J_0(\beta(\xi,x) \tilde{R})\frac{\sin(x\tilde{z})}{x}
    \end{split}
\end{equation}
and
\begin{equation}
    \begin{split}
        F_2(\xi, \tilde{b}) =
        & \frac{F_{\text{rel}}}{2}\xi\int^{\tilde{b}}_0 d\tilde{R}\int^{\sqrt{\tilde{b}^2-\tilde{R}^2}}_{0}d\tilde{z}
        \\& \times \left[\int^{\sqrt{4-\xi}}_0 dx J_0(\beta(\xi,x) \tilde{R})\frac{\sin(x \tilde{z})}{x}\right]
        \\& \times\left[\int^{\sqrt{4-\xi}}_0 dx J_0(\beta(\xi,x) \tilde{R})\cos(x \tilde{z})\right].
    \end{split}
\end{equation}
Here, 
\begin{equation}
    \beta(\xi,x) = \sqrt{\frac{-2x^2-\xi+\sqrt{16x^2+\xi^2}}{2}},
\end{equation}
\begin{equation}
    F_{\text{rel}} = 4\pi\bar{\rho}\left(\frac{G M}{v_{\text{rel}}}\right)^2,
\end{equation}
and $\tilde{b}$ is a cutoff.  We need to impose a cutoff otherwise the integral diverges.  As $\xi$ goes to zero, the expression correctly approaches that in the non-interacting limit.

\begin{figure}
    \centering
    \includegraphics[width=.48\textwidth]{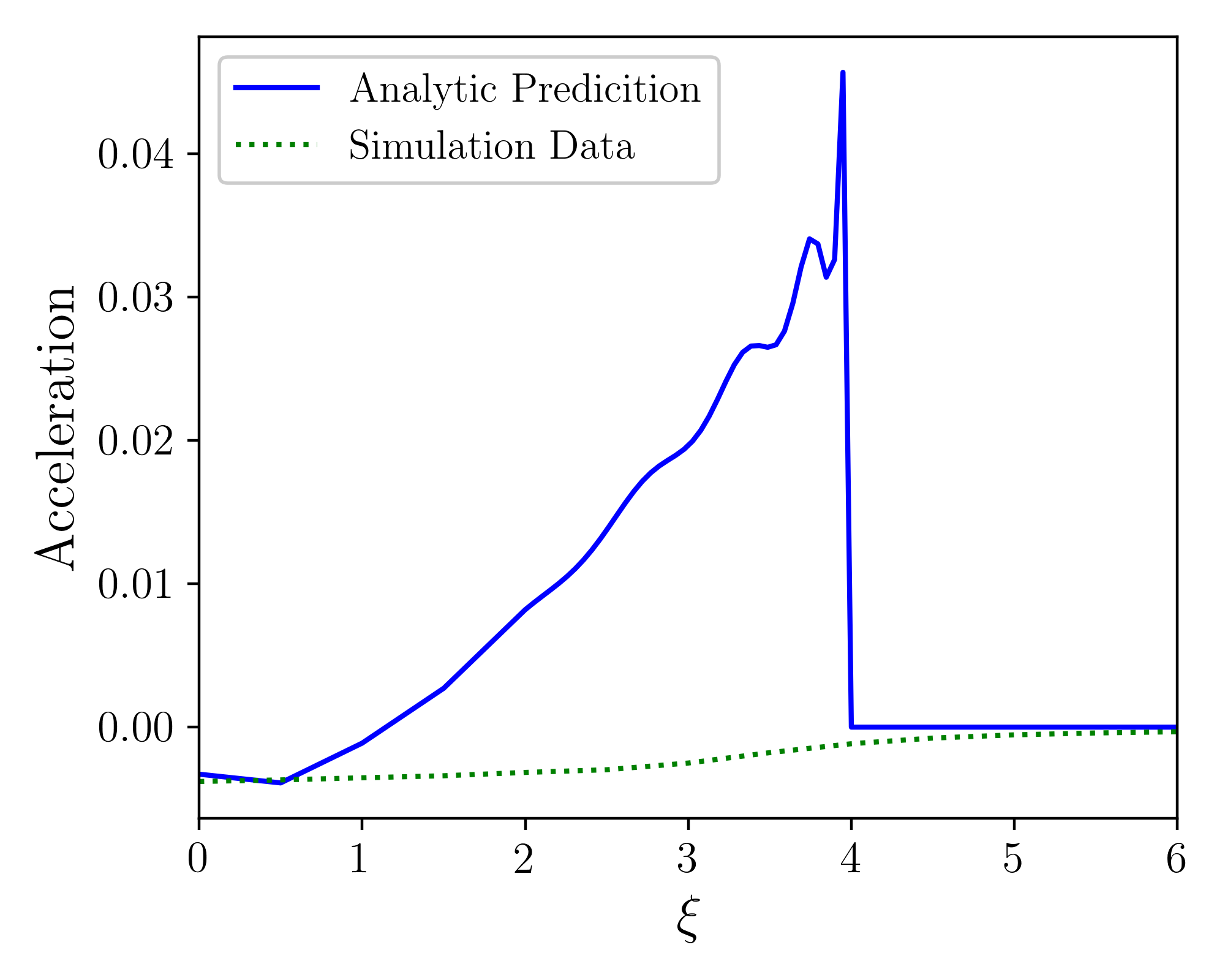}
    \caption{%
        The predicted acceleration due to dynamical friction from Eq.~\ref{DFresult} (solid blue) and the measured acceleration in the simulations (dotted green) for a cutoff of $\tilde{b} = 20$.  
        The two values are comparable for small $\xi$ and $\xi > 4$.
        However, there is a large discrepancy between then that grows as the value of $\xi$ approaches 4.
    }
    \label{fig:disagree}
\end{figure}

\section{Convergence Tests}
\label{sec:restest}

\begin{figure}
    \centering
    \includegraphics[width=.48\textwidth]{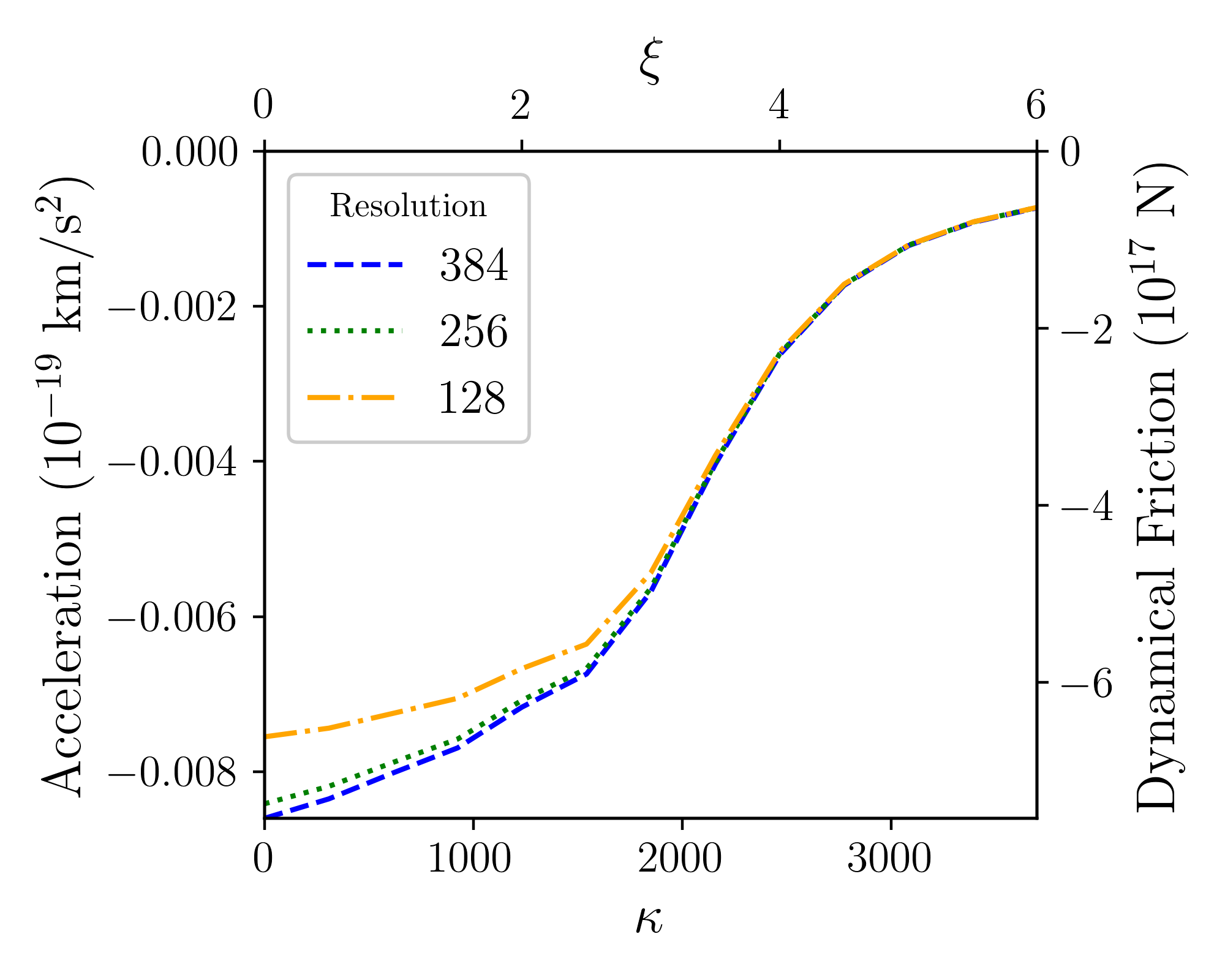}
    \caption{Measured dynamical friction force versus $\kappa$ for a point mass traveling through a uniform background at different resolutions.  
    Higher resolutions result in greater measured accelerations from dynamical friction.
    The dynamical friction forces converge when the resolution is $256^3$.
    The dynamical friction force measurements are closer together with more repulsive self-interactions.
    }
    \label{restest}
\end{figure}

\begin{figure}
    \centering
    \includegraphics[width=.48\textwidth]{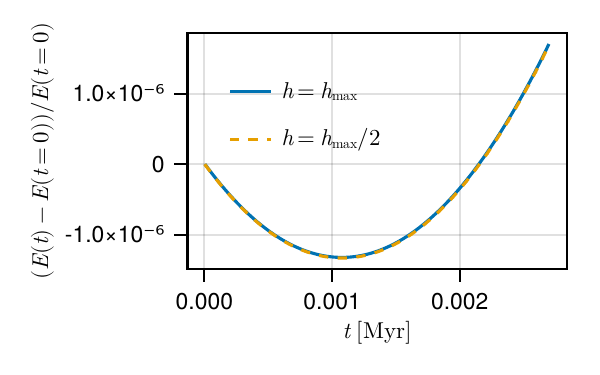}
    \caption{
        Relative change in total energy as a function of time, for a point mass traveling through a uniform background and $\lambda=0$
        The solid blue curve is for time step $h = h_{\mathrm{max}}$ defined in \cref{eq:timestep} and used in the other simulations.
        The dashed orange curve is for a smaller time step, $h = h_{\mathrm{max}} / 2$.
        In both cases, energy is conserved to 1 part in $10^6$, indicating that the time step is sufficiently small.
    }
    \label{fig:test_time_step}
\end{figure}

\begin{figure}[t!]
    \centering
    \includegraphics[width=.48\textwidth]{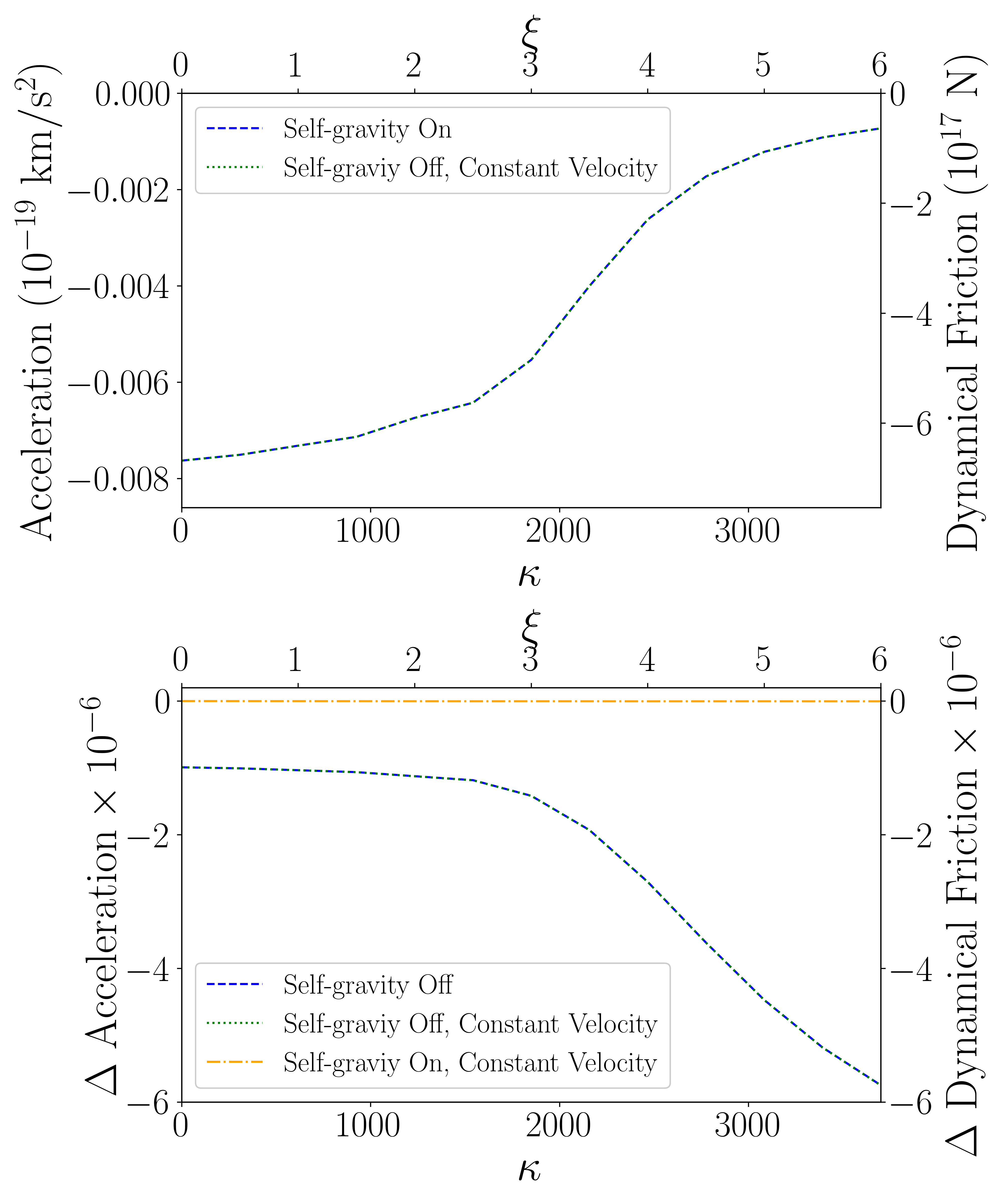}
    \caption{%
        Top: acceleration of a point particle as a function of $\kappa$ in cases where we assume the full dynamics as described in Sec.~\ref{sec:description} (dashed blue) versus no self-gravity of the ALP field or acceleration of the point particle, corresponding to the assumptions of our analytic calculation in Appendix~\ref{sec:analyticcalc} (dotted green).
        Bottom: relative difference of the acceleration as a function of time, as compared to the full dynamics of Sec.~\ref{sec:description} (dot-dashed orange), with self-gravity of the ALP field but no acceleration of the point particle (dashed blue), and with no self-gravity of the ALP field or acceleration of the point particle (dotted green).
        None of these differences significantly affect the measured acceleration.
    }
    \label{fig:approximation_consistency}
\end{figure}

We run additional simulations to verify the numerical convergence of our results presented in Sec.~\ref{sec:results}.
We find our fiducial results for the force of dynamical friction in a self-interacting ULDM background do not significantly depend on the resolution and box size.

To test the effects of these parameters, we set up simulations of a point particle traveling through a uniform ULDM background at different resolutions and box sizes.
In each case, the mass of the particle was $4.4\times10^{5}~M_{\odot}$, the initial velocity of the particle was $1.9\times10^{4}~\mathrm{km~s^{-1}}$, the background density  was $4.0\times10^{-6}~M_{\odot}~\mathrm{pc^{-3}}$, and $\kappa$ was from the set to be between $0$ and $3700$, yielding the same range of $\xi$ as in Sec.~\ref{sec:theoryresults}.

We ran three sets of simulations with a box size of $15~\mathrm{kpc}$ and resolutions of 128, 256, and 384.
The change in acceleration versus $\xi$ for these simulations are shown in Fig.~\ref{restest}.
The difference between the acceleration measurements is small between the sets run at 256 and 384 resolution.
All simulations in the main text use a resolution of 256, except where explicitly noted otherwise in this Appendix.

We also ran one set of simulations with a resolution of 256 and a box length $30~\mathrm{kpc}= 2 \times 15~\mathrm{kpc}$.
The results from this set of simulations were indistinguishable from those where the resolution was 128 and the box size was $15~\mathrm{kpc}$. The similarity of these results confirms that Jeans instability on the scale of the box does not affect our simulations.

We ran simulations to check energy conservation and that the time step of \cref{eq:timestep} is sufficiently small.
The results of this are displayed in \cref{fig:test_time_step}.
We find that energy is conserved to 1 part in $10^6$.
Furthermore, we find that this level of energy conservation is not changed when the time step is halved, indicating that the time step used in our other simulations is sufficiently small.

Finally, we ran simulations to check the assumptions of the analytic calculation in Appendix~\ref{sec:analyticcalc}.
This calculation makes two approximations that differ from our fiducial simulations.
First, it assumes that the velocity of the point particle remains constant.
This is valid in the limit that the mass contained in the gravitational wake is much smaller than that of the point particle. 
In addition, the analytic calculation assumes that the self-gravity of the ALP field does not contribute to the growth of the wake.
We checked the validity of these assumptions by running simulations in an alternate version of the code in which they are enforced and comparing the output to that from the unmodified code.
Fig.~\ref{fig:approximation_consistency} shows the results of these checks.
We found that the relative difference in measured acceleration is $\lesssim \mathcal{O}(10^{-3})$ across the range of $\xi$ values.
This is far too small to account for the discrepancy between our analytic predictions and fiducial simulation results, shown in Fig.~\ref{fig:disagree}. Thus, these assumptions are not the reason for the disagreement between our numerical and analytic results in certain regions of ULDM parameter space.

\end{document}